\documentclass[11pt, journal, letterpaper, oneside, onecolumn]{IEEEtran}

\usepackage[utf8]{inputenc}
\usepackage{graphicx} 
\graphicspath{ {Images/} }
\usepackage{epsfig}   
\usepackage{mathptmx} 
\usepackage{times}    
\usepackage{amsmath}  
\usepackage{amssymb}  
\usepackage{units}    
\usepackage{nicefrac}    
\usepackage{booktabs} 
\usepackage{cite}     
\usepackage{newtxtext, newtxmath} 
\usepackage{hyperref} 
\usepackage{float}    
\usepackage{color}    
\usepackage{subfig}   

\newtheorem{lemma}{Lemma}

\begin{document}
	\title{Robust Control of the Sit-to-Stand Movement\\ for a Powered Lower Limb Orthosis}
	\author{Octavio~Narvaez-Aroche$^1$~(ocnaar@berkeley.edu), Pierre-Jean~Meyer$^2$~(pjmeyer@berkeley.edu),\\ Stephen~Tu$^2$~(stephent@berkeley.edu),
		Andrew~Packard$^1$~(apackard@berkeley.edu),\\ and Murat~Arcak$^2$~(arcak@berkeley.edu)
		\thanks{$^1$Berkeley Center for Control and Identification, Department of Mechanical Engineering, University of California, Berkeley, CA, 94720.}%
		\thanks{$^2$Department of Electrical Engineering and Computer Sciences, University of California, Berkeley, CA, 94720 USA.}
	}
	\maketitle
	
	
	\begin{abstract}
		
		The sit-to-stand movement is a key feature for wide adoption of powered lower limb orthoses for patients with complete paraplegia. In this paper we study the control of the ascending phase of the sit-to-stand movement for a minimally actuated powered lower limb orthosis at the hips. First, we generate a pool of finite horizon Linear Quadratic Regulator feedback gains, designed under the assumption that we can control not only the torque at the hips but also the loads at the shoulders that in reality are applied by the user. Next we conduct reachability analysis to define a performance metric measuring the robustness of each controller against parameter uncertainty, and choose the best controller from the pool with respect to this metric. Then, we replace the presumed shoulder control with an Iterative Learning Control algorithm as a substitute for human experiments. Indeed this algorithm obtains torque and forces at the shoulders that result in successful simulations of the sit-to-stand movement, regardless of parameter uncertainty and factors deliberately introduced to hinder learning. Thus it is reasonable to expect that the superior cognitive skills of real users will enable them to cooperate with the hip torque controller through training.
		
	\end{abstract}
	
	
%
	\IEEEpeerreviewmaketitle
	
	\section{Introduction}
	
	Powered Lower Limb Orthoses (PLLOs) are medical devices worn in parallel to the legs to assist standing and/or walking. State of the art PLLOs for people with paraplegia ($ \approx 114,000 $ individuals in the USA \cite{NSCISC2018}) are commercially known as medical exoskeletons. Their users must have healthy enough skeletal, cardiovascular, vestibular, and visual systems to tolerate standing, as well as mobility in hands, arms, and shoulders to interact with the ground by means of crutches. The majority of exoskeletons are equipped with actuation at the hips and knees \cite{Cyberdyne2018,EksoBionics2018,ParkerHannifinCorp2018,ReWalkRobotics2018,RokiRobotics2018}, but the most affordable \cite{SuitX2018} uses a minimally actuated architecture where torque is exclusively applied at the hips \cite{McKinley2014}. In addition gait cycles on level ground with this design look more natural than those of its competitors. However, it is more difficult for users with complete paraplegia to perform the sit-to-stand (STS) movement, which is the sequence of actions for rising from a chair.
	\begin{figure}
		\begin{centering}
			\includegraphics[width=6.5cm]{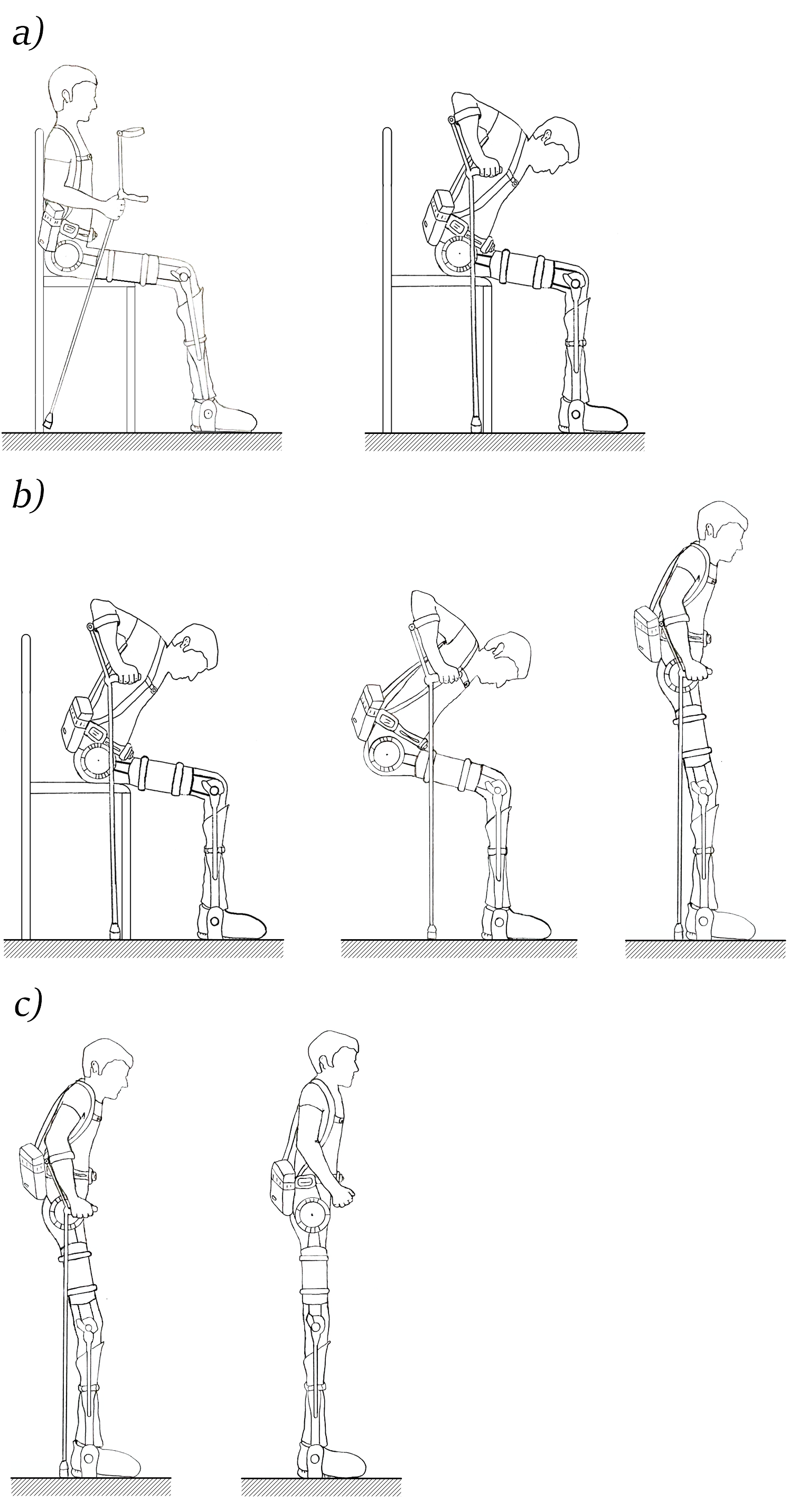}
			\par\end{centering}
		\caption{Phases of a dynamic sit-to-stand movement for a powered lower limb orthosis. a) Preparation. b) Ascension. c) Stabilization.\label{fig:STSPhases}}
	\end{figure}
	
	The STS movement is executed in three distinctive phases: preparation, ascension, and stabilization, as illustrated in Figure \ref{fig:STSPhases}. The ascension phase (Figure \ref{fig:STSPhases}b) starts at seat-off and ends when the links of the shanks and thighs segments almost align with the vertical, and the torso has a slight forward tilt, with all angular velocities close to zero in order to facilitate stabilization about the standing position. It is the most challenging phase because of the greater ranges of joint motion, torques, and forces involved. It also requires precise coordination between the actuators of the PLLO and the loads applied by the upper limbs of the user to avoid sit-back or step failures \cite{Eby2006}.
	
	This paper makes two main contributions. First, it provides a performance metric to quantify the robustness against parameter uncertainty of a controller for the ascension phase of a PLLO. Second, it assesses through numerical simulation if a controller would be suitable for clinical implementation, using an Iterative Learning Control (ILC) algorithm as a proxy for the human loads at the shoulders.
	
	Section \ref{sec:ControllerDesign} reviews the Euler-Lagrange equations of the three-link robot model of the system, and our motion plan for the PLLO, which is based on the desired angular position and velocity of the links for the thigh segment, and the kinematics of the center of mass (CoM). It also derives finite horizon LQR controllers to track the reference trajectories under the provisional assumption that the controller has authority not only over the torque of the actuators at the hips of the PLLO, but also over the torque and forces at the shoulders of the user.
	
	In Section \ref{sec:RobustPerformance} we select from a pool of finite horizon LQR controllers the one that optimizes a performance metric that measures robustness against parameter uncertainty. This metric is defined using a suitable reachability analysis when the parameters vary within given intervals. 
	
	
	Since the PLLO can only drive the actuators at the hips, in Section \ref{sec:ILC} we propose an ILC algorithm to simulate the loads at the shoulders that would be applied by a user, when being trained to perform the ascending phase of the STS movement in closed-loop with the optimal controller. To avoid identification experiments that would expose the user to non validated controllers, we tune the internal gains of the ILC with a reinforcement learning approach. We confirm that this simple proxy for the user achieves successful STS movements after a reasonable number of iterations despite considerable weight fluctuations and factors hindering learning.
	
	Companies producing PLLOs for people with complete paraplegia are moving towards stand-alone mobility solutions that can be operated outside of rehabilitation centres, and without the supervision of a specially trained physical therapist. This calls for extensive clinical trials for certifying the safety and feasibility of their designs to stand up and walk under a wide variety of conditions, as was done in \cite{Baunsgaard2018,Zeilig2012} to certify the potential benefits on gait function and balance. Even though our simulations cannot replace such tests, they can be valuable tools for improving both the mechanical design and control strategies of the devices prior to a comprehensive training protocol for the STS movement. 
	
	\paragraph*{Notation}
	Coordinate aligned boxes play an important role in this study. For $a,b\in\mathbb{R}^n$ we use the notation $a \leq b$ to mean $a_i \leq b_i$ $\forall i$, define an interval of $\mathbb{R}^n$ as $\left[a,b \right]:=\{\xi \in\mathbb{R}^n | a \leq \xi \leq b\}\subseteq\mathbb{R}^n$, and compute its volume $\mathrm{vol}([a,b])\in\mathbb{R}$ as: \[\mathrm{vol}([a,b])=\prod_{i\in\{1,\dots,n\}}(b_{i}-a_{i}).\]For matrices $ \Lambda,\, \underline{\Lambda}, \overline{\Lambda}  \in \mathbb{R}^{n \times m} $ we write $ \Lambda \in \left[ \underline{\Lambda}, \, \overline{\Lambda} \right] $ if $ \Lambda_{ij} \in \left[\underline{\Lambda}_{ij}, \, \overline{\Lambda}_{ij} \right] $ $ \forall (i, \, j) \in \{ 1, \ldots, n \} \times \{ 1, \ldots, m \} $. The center of the interval matrix $ \left[ \underline{\Lambda}, \, \overline{\Lambda} \right] $ is represented by $ \hat{\Lambda} $.
	
	\paragraph*{Acronyms}
	\begin{description}
		\item [{STS}] sit-to-stand.
		\item [{PLLO}] powered lower limb orthosis. 
		\item [{CoM}] center of mass.
		\item [{LQR}] Linear Quadratic Regulator.
		\item [{ILC}] Iterative Learning Control.
	\end{description}
	
	\section{Design of a Tracking Controller for the Powered Lower Limb Orthosis} \label{sec:ControllerDesign}
	
	This section describes the three-link robot model used for motion planning, control design, and reachability analysis. The contents of this section and Section \ref{subsec:Reachability} that follows were published in \cite{Narvaez-Aroche2018} and \cite{Narvaez-Aroche2018a}. They are included here for self-containment of the paper.
	
	\subsection{Model of the Powered Lower Limb Orthosis and its User}
	
\begin{figure}[b]
		\begin{centering}
			\includegraphics[width=5.5cm]{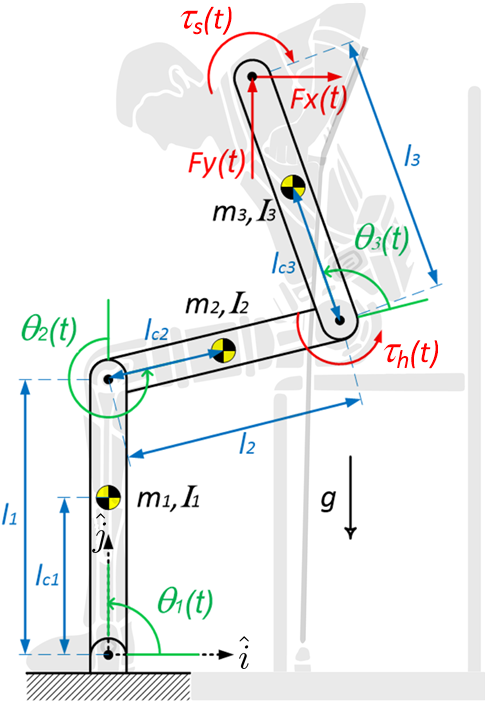}
			\par\end{centering}
		\caption{Three-link planar robot for modeling a powered lower limb orthosis (PLLO) during a sit-to-stand (STS) movement.\label{fig:Robot}}
	\end{figure}

	Assuming sagittal symmetry, no movement of the head relative to the torso, and that the feet are fixed to the ground, we model the user, crutches and PLLO as a three-link planar robot with revolute joints coaxial to the ankles, knees and hips, as shown in Figure~\ref{fig:Robot}. $\theta_{1}$ is the angular position of link 1 (shanks) measured from the horizontal, $\theta_{2}$ is the angular position of link 2 (thighs) relative to link 1, and $\theta_{3}$ is the angular position of link 3 (torso) relative to link 2. The system parameters are the masses of the links $m_1$, $m_2$, and $m_3$; the moments of inertia about their respective CoMs $I_{1}$, $I_{2}$, and $I_{3}$; their lengths $l_1$, $l_2$, and $l_3$; and the distances of their CoMs from the joints $l_{\scriptstyle{\mathrm{c}_1}}$, $l_{\scriptstyle{\mathrm{c}_2}}$, and $l_{\scriptstyle{\mathrm{c}_3}}$. The actuators of the orthosis exert torque $\tau_{h}$ about the hips; while torque $\tau_{s}$, horizontal force $F_{x}$ and vertical force $F_{y}$ capture the inertial and gravitational forces of the arms and loads applied on the shoulders of the user. There is no actuation at the knees.
	In terms of the joint angles vector $ \theta:=\left[\theta_{1};\;\; \theta_{2};\;\;\theta_{3}\right] $, input $u:=\left[\tau_{h};\;\; \tau_{s};\;\; F_{x};\;\; F_{y}\right]$, and parameter \[	p:=\left[m_{1};\;\;m_{2};\;\;m_{3};\;\;I_{1};\;\;I_{2};\;\;I_{3};\;\;l_{1};\;\;l_{2};\;\;l_{3};\;\;l_{\scriptstyle{\mathrm{c}_1}};\;\;l_{\scriptstyle{\mathrm{c}_2}};\;\;l_{\scriptstyle{\mathrm{c}_3}}\right],
	\]the Euler-Lagrange equations of the three-link planar robot in Figure \ref{fig:Robot} can be written as \begin{equation}
	M\left(\theta\left(t\right),p\right)\ddot{\theta}\left(t\right)+F\left(\theta\left(t\right),\dot{\theta}\left(t\right),p\right)=A_{\tau}\left(\theta\left(t\right),p\right)u\left(t\right), \label{eq:EulerLagrange}
	\end{equation}where $M\left(\theta,p\right)\in\mathbb{R}^{3\times3}$, $M\left(\theta,p\right)\succ0$ is the symmetric mass matrix of the system, $F\left(\theta,\dot{\theta},p\right)\in\mathbb{R}^{3}$ is the vector of energy contributions due to the acceleration of gravity and Coriolis forces, and $A_{\tau}\left(\theta,p\right)\in\mathbb{R}^{3\times4}$ is the generalized force matrix. Their entries are shown explicitly in Appendix \ref{appendix:EulerLagrange}. 
	
	Three rigid link dynamic models like this have been used to accurately describe the STS movement of human subjects \cite{Eby2006}.
	
	\subsection{Sit-To-Stand Motion Planning}
	
	Biomechanical studies measure the kinematics of the CoM of the human body instead of joint angles to classify and assess dynamic balance of the STS movement \cite{Fujimoto2012}. Therefore, considering $\theta_{2}$, and the position coordinates of the CoM of the three-link planar robot in its inertial frame $\left(x_{\scriptscriptstyle{\mathrm{CoM}}},y_{\scriptscriptstyle{\mathrm{CoM}}}\right)$, we define $ z:=\left[\theta_{2}; x_{\scriptscriptstyle{\mathrm{CoM}}}; y_{\scriptscriptstyle{\mathrm{CoM}}}\right]$ and plan the STS motion over the finite time horizon $t\in\left[t_{0},t_{f}\right]$ with reference trajectories \begin{align}\begin{split}
	\hat{\theta}_{2}\left(t\right) & =  \hat{\theta}_{2}\left(t_{0}\right)+\left(\hat{\theta}_{2}\left(t_{f}\right)-\hat{\theta}_{2}\left(t_{0}\right)\right)\Theta\left(t,t_{f}\right),\\
	\hat{x}_{\scriptscriptstyle{\mathrm{CoM}}}\left(t\right)    & =  \hat{x}_{\scriptscriptstyle{\mathrm{CoM}}}\left(t_{0}\right)+\left(\hat{x}_{\scriptscriptstyle{\mathrm{CoM}}}\left(t_{f}\right)-\hat{x}_{\scriptscriptstyle{\mathrm{CoM}}}\left(t_{0}\right)\right)\Theta\left(t,t_{f}\right),\\
	\hat{y}_{\scriptscriptstyle{\mathrm{CoM}}}\left(t\right)    & =  \hat{y}_{\scriptscriptstyle{\mathrm{CoM}}}\left(t_{0}\right)+\left(\hat{y}_{\scriptscriptstyle{\mathrm{CoM}}}\left(t_{f}\right)-\hat{y}_{\scriptscriptstyle{\mathrm{CoM}}}\left(t_{0}\right)\right)\Theta\left(t,t_{f}\right).
	\end{split}\label{eq:MotionPlanning}\end{align}For a rest-to-rest maneuver from $ \hat{z}\left(t_0\right) $ to $ \hat{z}\left(t_f\right) $, define $ \Theta \left(t,t_{f}\right):=-2\frac{t^{3}}{t_{f}^{3}}+3\frac{t^{2}}{t_{f}^{2}} $; which is the only cubic polynomial satisfying $ \dot{\Theta}\left(t_{0},t_{f}\right)=\dot{\Theta}\left(t_{f},t_{f}\right)=0 $, $ \Theta\left(t_{0},t_{f}\right)=0 $, and $ \Theta \left(t_{f},t_{f}\right) = 1 $.
	
	Relying on kinematic equations, it was shown in \cite{Narvaez-Aroche2017} that for feasible and realistic STS movements excluding the vertical position ($ \theta_{1}=\nicefrac{\pi}{2} $, $ \theta_{2}=\theta_{3}=0 $), a transformation of the form\begin{equation}
	\left[
	\hat{\theta}\left(t\right); \;\; \dot{\hat{\theta}}\left(t\right); \;\; \ddot{\hat{\theta}}\left(t\right)\right]=h\left(\hat{z}\left(t\right),\dot{\hat{z}}\left(t\right),\ddot{\hat{z}}\left(t\right),\hat{p}\right)
	\label{eq:z2theta}\end{equation}exists; so that the reference trajectories for the ascending phase in the $z$ space from \eqref{eq:MotionPlanning} can be mapped into the $\theta$ space with the nominal value of the parameter $\hat{p}$. The derivation of this mapping is included in Appendix \ref{appendix:z2thetaTransformation}.
	
	We use the computed torque method \cite{Slotine1991} to get reference trajectories $ \hat{u}\left(t\right) $. However, as the system of equations in (\ref{eq:EulerLagrange}) is underdetermined, at every $t\in\left[t_{0}, \, t_{f}\right]$ a control allocation problem \cite{Johansen2013} is solved to minimize the 2-norm of the input weighted by $ W_{u}\in\mathbb{R}^{4\times4} $ in the presence of box constraints $ \underline{u}, \, \overline{u}\in~\mathbb{R}^{4} $:\begin{align}
	\hat{u}\left(t\right)= & \:\underset{\xi\in\mathbb{R}^4}{\arg\min} \quad \frac{1}{2}\left\Vert W_{u}\:\xi\right\Vert _{2}^{2} \label{eq:Allocation} \\
	& \text{subject to} \nonumber \\ & \quad A_{\tau}\left(\hat{\theta}\left(t\right),\hat{p}\right)\xi=M\left(\hat{\theta}\left(t\right),\hat{p}\right)\ddot{\hat{\theta}}\left(t\right)+F\left(\hat{\theta}\left(t\right),\dot{\hat{\theta}}\left(t\right),\hat{p}\right)\nonumber\\
	& \quad \quad \quad \quad \underline{u}\leq\xi\leq \overline{u}.\nonumber
	\end{align}We take $W_u:=\operatorname{diag}\left(\left[ 1, 1, 10, 1\right]\right)$ to ensure that the contributions from $\hat{\tau}_{h}\left(t\right)$, $\hat{\tau}_{s}\left(t\right)$ and $\hat{F}_{y}\left(t\right)$ outweigh $\hat{F}_{x}\left(t\right)$. To account for the limits of torque for the actuators at the hips, as well as the ones for the loads applied by the user at the shoulders, we set $ \underline{u}:= \left[-200\,[\mathrm{N \cdot m}]; \, -175\,[\mathrm{N \cdot m}]; \, -40\,[\mathrm{N}]; \, 0 [\mathrm{N}] \right] $ and $ \overline{u}:=~\left[200 \, [\mathrm{N \cdot m}]; \, 50 \, [\mathrm{N \cdot m}]; \, 40 \, [\mathrm{N}]; \, 650 \, [\mathrm{N}] \right] $. Different values for $ W_{u} $, $ \underline{u} $, and $ \overline{u} $ might lead to different $ \hat{u}\left(t\right) $ for the same $\hat{\theta}(t)$, $\dot{\hat{\theta}}(t)$, $\ddot{\hat{\theta}}(t)$, and $\hat{p}$. Since a subject pushes the crutches down to propel upwards during a STS movement, the constraint $ F_{y}\left(t\right)\geq 0 $ must always be imposed.
	
	\subsection{Finite Time Horizon LQR Controller}
	\label{subsec:LQR Design}
	
	Defining the state $ x\in\mathbb{R}^6 $ as $ x:=\left[\,\theta; \:\dot{\theta}\right] $, we first note from \eqref{eq:EulerLagrange} that the dynamics of the three-link planar robot are\begin{align}
	\dot{x}\left(t\right) & = \left[\begin{array}{c}
	\dot{\theta}\left(t\right)\\
	M^{-1}\left(\theta\left(t\right),p\right)\left(A_{\tau}\left(\theta\left(t\right),p\right)u\left(t\right)-F\left(\theta\left(t\right),\dot{\theta}\left(t\right),p\right)\right)
	\end{array}\right] \nonumber \\
	& =: f\left(x\left(t\right),p,u\left(t\right)\right). \label{eq:ThreeLinkRobotModel}
	\end{align}
	Next we linearize \eqref{eq:ThreeLinkRobotModel} to design a finite horizon LQR controller for tracking the reference state trajectory $\hat{x}\left(t\right):=\left[\hat{\theta}\left(t\right),\dot{\hat{\theta}}\left(t\right)\right]$ obtained from~\eqref{eq:MotionPlanning} and~\eqref{eq:z2theta}. The state deviation variable $\delta_x\left(t\right):=x\left(t\right)-\hat{x}\left(t\right)$ satisfies \[
	\dot{\delta}_{x}\left(t\right)=f\left(x\left(t\right),p,u\left(t\right)\right)-f\left(\hat{x}\left(t\right),\hat{p},\hat{u}\left(t\right)\right),
	\]which can be approximated with a first order Taylor series expansion of $f\left(x\left(t\right),p,u\left(t\right)\right)$ about $\hat{x}\left(t\right)$, $\hat{p}$ and $\hat{u}\left(t\right)$:\begin{align}
	\dot{\delta}_{x}\left(t\right) \approx& \left.\frac{\partial f\left(x,p,u\right)}{\partial x}\right|_{\scriptsize {\begin{array}{l}
			x=\hat{x}\left(t\right)\\
			p=\hat{p}\\
			u=\hat{u}\left(t\right)
			\end{array}}}\left(x\left(t\right)-\hat{x}\left(t\right)\right)
+\left.\frac{\partial f\left(x,p,u\right)}{\partial p}\right|_{\scriptsize {\begin{array}{l}
			x=\hat{x}\left(t\right)\\
			p=\hat{p}\\
			u=\hat{u}\left(t\right)
			\end{array}}}\left(p-\hat{p}\right)
+\left.\frac{\partial f\left(x,p,u\right)}{\partial u}\right|_{\scriptsize {\begin{array}{l}
			x=\hat{x}\left(t\right)\\
			p=\hat{p}\\
			u=\hat{u}\left(t\right)
			\end{array}}}\left(u\left(t\right)-\hat{u}\left(t\right)\right)\nonumber\\
	=& A\left(t\right)\delta_{x}\left(t\right)+B_{1}\left(t\right)\delta_{p}+B_{2}\left(t\right)\delta_{u}\left(t\right).\label{eq:LTV}
	\end{align}
	From \cite{Athans1966}, for unconstrained $ \delta_{u}\left(t\right) $, symmetric matrices $ Q, \, S\succeq~0 $, and $ R\succ0 $, the optimal control of the stabilizable linear time varying system in (\ref{eq:LTV}) with quadratic cost \begin{align*}
	J_{\scriptscriptstyle{\mathrm{LQR}}}=&\frac{1}{2}\delta_{x}^{\top}\left(t_{f}\right)S\delta_{x}\left(t_{f}\right)+\frac{1}{2}\int_{t_{0}}^{t_{f}}\left( \delta_{x}^{\top}\left(t\right)Q\delta_{x}\left(t\right)+\delta_{u}^{\top}\left(t\right)R\delta_{u}\left(t\right)\right) \, dt    
	\end{align*}exists, and is unique, given by the time varying formula\begin{align} \begin{split}
	\delta_{u}\left(t\right) & = -R^{-1}B_{2}^{\top}\left(t\right)P\left(t\right)\delta_{x}\left(t\right)\\
	& =: -K_{\scriptscriptstyle{\mathrm{LQR}}}\left(t\right)\delta_{x}\left(t\right). \label{eq:LQRgain}
	\end{split} \end{align}With the boundary condition $P\left(t_{f}\right)=S$, $P\left(t\right)$ is the solution of the Riccati matrix differential equation \begin{align}
	\dot{P}\left(t\right)=&-P\left(t\right)A\left(t\right)-A^{\top}\left(t\right)P\left(t\right)+P\left(t\right)B_{2}\left(t\right)R^{-1}B_{2}^{\top}\left(t\right)P\left(t\right)-Q.\label{eq:Ric}
	\end{align}
	The nonlinear dynamics of the three-link robot under state feedback control with the time varying matrix gain $ K_{\scriptscriptstyle{\mathrm{LQR}}}\left(t\right) \in \mathbb{R}^{4 \times 6}$, become\begin{align}
	\dot{x}\left(t\right) & =f\left(x\left(t\right),p,\hat{u}\left(t\right)-K_{\scriptscriptstyle{\mathrm{LQR}}}\left(t\right)\left(x\left(t\right)-\hat{x}\left(t\right)\right)\right)\nonumber\\
	&=: \varphi \left(t,x,p\right). \label{eq:Nonlinear}
	\end{align}
	
	\section{Robust Performance Metric Under Parameter Uncertainty} \label{sec:RobustPerformance}
	
	There is no guarantee that a choice of $Q$, $R$, $S$ in the design of the finite time horizon LQR controller of the previous section will achieve a safe STS movement in the presence of parameter uncertainty, i.e., when $ p $ is an unknown constant lying within an interval, due to manufacturing variability of the links of the PLLO and weight fluctuations of its user. To properly evaluate the robustness, we define a performance metric for assessing the worst-case deviations of the state $ x(t) $, output $ y(t):=[x_{\scriptscriptstyle{\mathrm{CoM}}}(t);y_{\scriptscriptstyle{\mathrm{CoM}}}(t);\dot x_{\scriptscriptstyle{\mathrm{CoM}}}(t);\dot y_{\scriptscriptstyle{\mathrm{CoM}}}(t)] $, and input $ u(t) $ from their desired trajectories $ \hat{x}(t) $, $ \hat{y}(t) $, and $ \hat{u}(t) $, based on over-approximations of their reachable sets at particular instants of time. We then identify the controller that optimizes the performance metric over a pool of candidates, as the most suitable for implementation.
	
	Since ascension starts from rest, with the shanks and torso segments parallel to the vertical, and the thighs segment parallel to the horizontal, we set \[ x \left(t_{0}\right)=\left[90[^\circ];\; -90[^\circ];\; 90[^\circ];\; 0[^\circ/\mathrm{s}];\; 0[^\circ/\mathrm{s}];\; 0[^\circ/\mathrm{s}] \right]. \]With nominal parameter\begin{align*}
	\hat{p} := [& 9.68\left[ \mathrm{kg} \right]; \; 12.59\left[ \mathrm{kg} \right];\; 44.57\left[ \mathrm{kg} \right]; 1.16\left[ \mathrm{kg \cdot m^2} \right];\; 0.52\left[ \mathrm{kg \cdot m^2} \right];\; 2.56\left[ \mathrm{kg \cdot m^2} \right]; \ldots \\ 
	& 0.53\left[ \mathrm{m} \right]; \; 0.41\left[ \mathrm{m} \right]; \; 0.52\left[ \mathrm{m} \right]; 0.27 \left[ \mathrm{m} \right]; \; 0.21\left[ \mathrm{m} \right]; \; 0.26\left[ \mathrm{m} \right]],
	\end{align*}the corresponding initial position of the CoM of the three-link robot is $[\hat{x}_{\scriptscriptstyle{\mathrm{CoM}}}\left(t_{0}\right); \, \hat{y}_{\scriptscriptstyle{\mathrm{CoM}}}\left(t_{0}\right)]=[0.31; \, 0.67]$[m]. We calculate $\hat{x}\left(t\right)$ and $ \hat{u}(t) $ from \eqref{eq:z2theta} and \eqref{eq:Allocation} with $ t_{0}=0 $[s] and $ t_{f}=3.5 $[s], and a final configuration that places the CoM directly above the origin of the inertial frame with the values $ \hat{\theta}_{2}\left(t_{f}\right)=-5[^\circ] $, $ \hat{x}_{\scriptscriptstyle{\mathrm{CoM}}}\left(t_{f}\right)=0 $[m], and $\hat{y}_{\scriptscriptstyle{\mathrm{CoM}}}\left(t_{f}\right)=0.97$[m]. The desired output $ \hat{y}(t) $ is determined with the mapping $ \zeta: \mathbb{R}^6 \times \mathbb{R}^{12} \rightarrow \mathbb{R}^4 $ in Appendix \ref{appendix:CoMKinematics}, so that $ \hat{y}(t) = \zeta\left(\hat{x}(t),\hat{p}\right) $. 
	
	The numerical computation of the first order Taylor series expansion in \eqref{eq:LTV} gives the time varying matrices $ A(t) $, $ B_1(t) $ and $ B_2(t) $ required in~\eqref{eq:LQRgain}-\eqref{eq:Ric} to get $ K_{\scriptscriptstyle{\mathrm{LQR}}}\left(t\right) $ as a function of the weight matrices $ Q $, $ R $ and $ S $.
	
	\pagebreak
	It is assumed that the unknown parameter of the system lies within the interval $ \left[\underline p,\overline p \right]\subseteq\mathbb{R}^{12} $ in Table~\ref{tab:UncertainP}, which was calculated for a fluctuation of $ \pm 5 \% $ of the nominal weight of the user with anthropometric data from~\cite{Bartel2006}.
	\begin{table}[h]
		\caption{Bounds for the Parameter Uncertainty of the System $ [\underline{p}, \overline{p} ]$ \label{tab:UncertainP}}
		\centering{}
		\begin{tabular}{ccccc}
			\toprule
			Link & $m_{i}\:\left[\mathrm{kg}\right]$ & $I_{i}\:\left[ \mathrm{kg\cdot m^{2}} \right]$ & $l_{i}\:\left[ \mathrm{m} \right]$ & $l_{ci}\:\left[ \mathrm{m} \right]$\tabularnewline
			\midrule
			\midrule 
			1 & $\left[9.2, 10.2\right]$ & $\left[1.10, 1.21\right]$ & $\left[0.52, 0.54\right]$ & $\left[0.23, 0.30\right]$\tabularnewline
			\midrule 
			2 & $\left[11.2, 13.2\right]$ & $\left[0.49, 0.54\right]$ & $\left[0.39, 0.42\right]$ & $\left[0.17, 0.23\right]$\tabularnewline
			\midrule 
			3 & $\left[42.3, 46.8\right]$ & $\left[2.40, 2.65\right]$ & $\left[0.51, 0.53\right]$ & $\left[0.24, 0.28\right]$\tabularnewline
			\bottomrule
		\end{tabular}
	\end{table}
	
	\subsection{Sensitivity-based Reachability Analysis} \label{subsec:Reachability}
	
	Consider a continuous-time, time varying system $\dot{\eta}=g(t,\eta,\rho)$ with state $ \eta \in \mathbb{R}^{n_{\eta}} $, uncertain parameter $ \rho \in \left[ \underline{\rho}, \, \overline{ \rho }\right] \subseteq \mathbb{R}^{n_\rho} $, and continuously differentiable vector field $ g: \mathbb{R} \times \mathbb{R}^{n_\eta} \times \mathbb{R}^{n_\rho} \rightarrow \mathbb{R}^{n_\eta} $. Denoting the state reached by this system at time $ t \geq t_0 $ from fixed initial state $ \eta_0 $ as $ \Phi(t;t_0,\eta_0,\rho)$, we write the reachable set under parameter uncertainty as \[ \operatorname{Reach}(t,[\underline \rho,\overline \rho]):=\{\Phi(t;t_0,\eta_0,\rho)~|~\rho\in[\underline \rho,\overline \rho]\}\subseteq\mathbb{R}^{n_\eta},\]and the sensitivity function of the state trajectories with respect to the parameter as \[S(t;t_0,\eta_0,\rho):=\frac{\partial\Phi(t;t_0,\eta_0,\rho)}{\partial \rho} \in \mathbb{R}^{n_\eta \times n_\rho}.\]We take the following lemma from \cite{meyer2018sampled}.
	
	\begin{lemma} \label{PJLemma}
		Assume that there exist $ \underline{\mathcal{S}},\, \overline{\mathcal{S}}:[t_0,+\infty) \rightarrow \mathbb{R}^{ n_\eta \times n_\rho } $ such that $S(t;t_0,\eta_0,\rho) \in \left[ \underline{\mathcal{S}}(t),\,\overline{\mathcal{S}}(t) \right]$ for all $t \geq t_0$, and $ \rho \in \left[ \underline{\rho}, \, \overline{\rho} \right] $. Let the functions  $ \underline{r}, \, \overline{r}:[t_0,+\infty) \rightarrow \mathbb{R}^{n_\eta} $ be defined as \begin{align}
		\underline{r}_i(t)&:=\Phi_i(t;t_0,\eta_0,\underline{\pi}^i(t))-d^i(t)(\underline{\pi}^i(t)-\overline{\pi}^i(t)), \nonumber\\
		\overline{r}_i(t)&:=\Phi_i(t;t_0,\eta_0,\overline{\pi}^i(t))+d^i(t)(\underline{\pi}^i(t)-\overline{\pi}^i(t)), \label{eq:r}
		\end{align}where the $j^{th}$ elements of the parameter values $ \underline{\pi}^{i}(t), \, \overline{\pi}^{i}(t) \in \left[ \underline{\rho}, \overline{\rho} \right] $ and row vector $ d^{i}(t) \in \mathbb{R}^{n_\rho} $ are determined according to the sign of the entries of the center of the interval matrix $ \left[ \underline{\mathcal{S}}(t),\,\overline{\mathcal{S}}(t) \right] $ as \[ \begin{array}{cc}
		\left. \begin{array}{l}
		\underline{\pi}^i_j(t):=\underline{\rho}_j, \; \overline{\pi}^i_j(t):=\overline{\rho}_j, \\ d^i_j(t):=\min \left(0, \, \underline{\mathcal{S}}_{ij}(t) \right)
		\end{array} \right\} \text{if} \; \hat{\mathcal{S}}_{ij}(t)\geq 0; \\  \\ \left.  \begin{array}{l}
		\underline{\pi}^i_j(t):=\overline{\rho}_j,\;\overline{\pi}^i_j(t):=\underline{\rho}_j, \\ d^i_j(t):=\max \left(0, \, \overline{\mathcal{S}}_{ij}(t) \right)
		\end{array} \right\} \text{if} \; \hat{\mathcal{S}}_{ij}(t)< 0.
		\end{array} \] Then $ \left[ \underline{r}(t), \, \overline{r} (t) \right] $ is an interval over-approximation of the reachable set of states at time $ t \geq t_{0} $, so that \[\operatorname{Reach}\left(t,\left[\underline \rho,\overline \rho \right] \right) \subseteq \left[ \underline{r}(t), \, \overline{r} (t) \right].\]
	\end{lemma}
	
	Given fixed $ x_0 := x(t_0) $ for the configuration of the links at seat-off, the sensitivity function $ S^{x}(t;t_0,x_0,p) $ for the state trajectory $ \Phi^{x}(t;t_0,x_0,p) $ of \eqref{eq:Nonlinear} satisfies the differential equation \begin{align}
	\dot S^x(t;t_0,x_0,p) &=\left.\frac{\partial \varphi(t,x,p)}{\partial x}\right|_{x=\Phi^{x}(t;t_0,x_0,p)} S^x(t;t_0,x_0,p) \quad + \left.\frac{\partial \varphi(t,x,p)}{\partial p}\right|_{x=\Phi^{x}(t;t_0,x_0,p)},\label{eq:SensitivityXMDE}
	\end{align}over $ t \in \left[ t_0, t_f \right] $, and zero initial condition $ S^{x}\left( t; t_0,x_0,p\right)=0_{6 \times 12} $ \cite{Khalil2001}. 
	
	The successors of $ y $ and $ u $ are determined from the static mappings of $ x $ and $ p $ that describe the kinematics of the CoM, and the control input with LQR state feedback:\begin{align*}
	\Phi^{y}(t;t_0,x_0,p):=& \zeta \left( \Phi^{x}\left(t;t_0,x_0,p \right), p \right), \nonumber\\
	\Phi^{u}(t;t_0,x_0,p):=& \hat{u}(t)-K_{\scriptscriptstyle{\mathrm{LQR}}}(t)\left(\Phi^{x}\left(t; t_0, x_0, p \right)-\hat x(t)\right).
	\end{align*}Therefore, their sensitivity functions can be obtained from the solution of \eqref{eq:SensitivityXMDE} as\begin{align}
	S^{y}\left(t;t_0,x_0,p\right) &:=\left.\frac{\partial\zeta\left(x,p\right)}{\partial x}\right|_{x=\Phi^{x}\left(t;t_0,x_0,p\right)}S^{x}\left(t;t_0,x_0,p\right) +\left.\frac{\partial\zeta\left(x,p\right)}{\partial p}\right|_{x=\Phi^{x}\left(t;t_0,x_0,p\right)} \in \mathbb{R}^{4 \times 12}, \nonumber \\
	S^{u}\left(t;t_0,x_0,p\right)&:=-K_{\scriptscriptstyle{\mathrm{LQR}}}\left(t\right)S^{x}\left(t;0,x_0,p\right) \in \mathbb{R}^{4 \times 12}. \label{eq:SensitivitiesYU}
	\end{align}
	
	We showed in \cite{Narvaez-Aroche2018a} that tight over-approximation functions for the reachable sets of the state $ \operatorname{Reach}^x\left(t,\left[\underline{p},\overline{p} \right] \right) \subseteq \left[ \underline{r}^x(t), \, \overline{r}^x (t) \right] $, output $ \operatorname{Reach}^y\left(t,\left[\underline{p},\overline{p} \right] \right) \subseteq \left[ \underline{r}^y(t), \, \overline{r}^y (t) \right] $, and input $ \operatorname{Reach}^u\left(t,\left[\underline{p},\overline{p} \right] \right) \subseteq \left[ \underline{r}^u(t), \, \overline{r}^u(t) \right] $, can be obtained from Lemma \ref{PJLemma} through a sampling approach. This consist in randomly drawing a set of 500 parameters $ \mathcal{P}_{b} \subset \left[ \underline{p}, \overline{p} \right] $ from a Latin Hypercube \cite{McKay1979}, and numerically solving~\eqref{eq:SensitivityXMDE} for each of them. The sensitivity bounds $ \left[ \underline{\mathcal{S}}^{x}(t),\,\overline{\mathcal{S}}^{x}(t) \right] $ are directly estimated by minimizing/maximizing the $ ij $ entries of the solutions for $ S^{x}(t;t_0,x_0,p) $ at time $ t $ for all $ p \in \mathcal{P}_{b} $. The estimates for $ \left[ \underline{\mathcal{S}}^{y}(t),\,\overline{\mathcal{S}}^{y}(t) \right] $, and $ \left[ \underline{\mathcal{S}}^{u}(t),\,\overline{\mathcal{S}}^{u}(t) \right] $ require first plugging the solutions sampled for the state sensitivity in \eqref{eq:SensitivitiesYU}, and then computing the extremal values for their respective entries.
	
	\subsection{Robust Performance Metric}
	\label{subsec:performance metric}
	
	For $ \left[a,\,b\right] \subseteq \mathbb{R}^n $, and $ c \in \mathbb{R}^n $, let $ \nu \left( \left[ a, \, b \right], c \right) \in \mathbb{R} $ be \begin{align*}
	\nu \left( \left[ a, \, b \right], c \right) &:= \prod_{ i=\{ 1, \ldots, n\} } \left| \frac{a_{i} + b_{i}}{2} - c_{i} \right|.
	\end{align*}
	To evaluate the worst-case performance of different controllers for tracking $ \hat{x}(t) $, $ \hat{u}(t) $, and $ \hat{y}(t) $, we propose the metric:
	\begin{align}
	J_{\scriptscriptstyle{\mathrm{P}}} := & \sum_{t\in T_{\mathrm{\scriptscriptstyle{P}}}} w^{\top}_{\scriptscriptstyle{\mathrm{v}}} \left[ \begin{array}{c} \\
	\mathrm{vol} \left( \left[ \underline{r}^{x}(t), \, \overline{r}^{x}(t) \right] \right)\\ \\
	\mathrm{vol} \left( \left[ \underline{r}^{y}(t), \, \overline{r}^{y}(t) \right] \right)\\ \\
	\mathrm{vol} \left( \left[ \underline{r}^{u}(t), \, \overline{r}^{u}(t) \right] \right)\\ \\
	\end{array} \right] + \sum_{t\in T_{\mathrm{\scriptscriptstyle{P}}}} w^{\top}_{\scriptscriptstyle{\mathrm{o}}} \left[ \begin{array}{c} \\
	\nu \left( \left[ \underline{r}^{x}(t), \, \overline{r}^{x}(t) \right], \,  \hat{x}(t) \right)\\ \\ 
	\nu \left( \left[ \underline{r}^{y}(t), \, \overline{r}^{y}(t) \right], \,  \hat{y}(t) \right)\\ \\
	\nu \left( \left[ \underline{r}^{u}(t), \, \overline{r}^{u}(t) \right], \,  \hat{u}(t) \right)\\ \\
	\end{array} \right],\label{eq:PerformanceMetric}
	\end{align}
	where $T_{\mathrm{\scriptscriptstyle{P}}} \subseteq \left[ t_0, \, t_f \right] $ is the set of time instants where the over-approximation functions are computed, $ w_{\scriptscriptstyle{\mathrm{v}}} \in \mathbb{R}^3 $ weighs the volumes enclosed by the intervals defined by such functions, and $ w_{\scriptscriptstyle{\mathrm{o}}} \in \mathbb{R}^3 $ weighs the volumes of the offsets between the center of the intervals and their reference trajectories. 
	
	To have a baseline value for $ J_{\mathrm{\scriptscriptstyle{P}}} $, we choose $ T_{\mathrm{\scriptscriptstyle{P}}}:=~\{0,0.875,1.75,2.625,3.5\} $[s], and compute\[ \begin{array}{ll}
	\sum_{t\in T_{\mathrm{\scriptscriptstyle{P}}}} \mathrm{vol} \left( \left[ \underline{r}^{x}(t), \, \overline{r}^{x}(t) \right] \right),& 
	\sum_{t\in T_{\mathrm{\scriptscriptstyle{P}}}} \nu \left( \left[ \underline{r}^{x}(t), \, \overline{r}^{x}(t) \right], \,  \hat{x}(t) \right),\\
	\sum_{t\in T_{\mathrm{\scriptscriptstyle{P}}}} \mathrm{vol} \left( \left[ \underline{r}^{y}(t), \, \overline{r}^{y}(t) \right] \right),& 
	\sum_{t\in T_{\mathrm{\scriptscriptstyle{P}}}} \nu \left( \left[ \underline{r}^{y}(t), \, \overline{r}^{y}(t) \right], \, \hat{y}(t) \right),\\ 
	\sum_{t\in T_{\mathrm{\scriptscriptstyle{P}}}} \mathrm{vol} \left( \left[ \underline{r}^{u}(t), \, \overline{r}^{u}(t) \right] \right),&
	\sum_{t\in T_{\mathrm{\scriptscriptstyle{P}}}} \nu \left( \left[ \underline{r}^{u}(t), \, \overline{r}^{u}(t) \right], \, \hat{u}(t) \right), \end{array}\]for the system in \eqref{eq:Nonlinear} under the action of the finite horizon LQR controller from \cite{Narvaez-Aroche2018}; which causes undesired variations of the loads at the shoulders \cite{Narvaez-Aroche2018a}. Using the reciprocals of these values, we set the weights in \eqref{eq:PerformanceMetric} to\begin{align*}
	w_{\scriptscriptstyle{\mathrm{v}}} &:= [ 6.98 \times 10^{7}; \; 9.67 \times 10^{-7}; \; 9.71 \times 10^{4}],\\
	w_{\scriptscriptstyle{\mathrm{o}}} &:= [ 1.85 \times 10^{18}; \; 7.24; \; 1.07 \times 10^{13} ],
	\end{align*}so that the performance metric for this baseline controller is $ J_{\mathrm{\scriptscriptstyle{P}}} = 6 $. The large difference in the order of magnitude of the weight entries is due to the units and dimensionality of the hypercubes from which they are calculated.
	
\begin{figure}
		\begin{centering}
			\includegraphics[width=8cm]{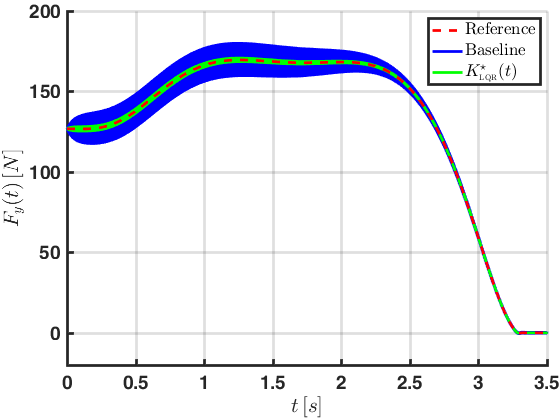}
			\par\end{centering}
		\caption{Comparison of the vertical force obtained for 500 randomly sampled parameter values, when the system is under the action of the baseline controller (blue), and under the optimal controller with respect to the performance metric $ J_{\mathrm{\scriptscriptstyle{P}}} $ (green). \label{fig:FyComparison}}
	\end{figure}

	Computing the over-approximation functions in \eqref{eq:PerformanceMetric} is too expensive to implement a derivative free optimization method, such as the one used in Section~\ref{sub:DFO} to tune the ILC gains. Hence here we opt for a brute force approach where we construct sets of $ 300 $ diagonal, positive definite matrices of LQR weight candidates $\mathcal{Q}\subset\mathbb{R}^{6\times6}$, $\mathcal{R}\subset\mathbb{R}^{4\times4}$, and $\mathcal{S}\subset\mathbb{R}^{6\times6}$. Their entries are randomly drawn from a Latin Hypercube of $ 300 $ samples on $ 16 $ variables, with the values for $ \mathcal{Q} $ and $ \mathcal{S} $ in $ \left(0,100 \right) $, and the ones for $ \mathcal{R} $ in $ \left(0,0.01\right) $. Each of the sampled triplets of weights are plugged into \eqref{eq:Ric}, which is solved with tools documented in \cite{Moore2015} to obtain their corresponding time-varying matrix gain $ K_{\scriptscriptstyle{\mathrm{LQR}}} \left( t \right) $ from~\eqref{eq:LQRgain}. Then the technique described in Section \ref{subsec:Reachability} is applied to find the over-approximation functions and calculate $ J_{\scriptscriptstyle{\mathrm{P}}} $ for all the controllers. The triplet of weight matrices \begin{equation*}
	Q^{\star},R^{\star},S^{\star}=\underset{Q\in\mathcal{Q},R\in\mathcal{R},S\in\mathcal{S}}{\arg\min}J_{\mathrm{\scriptscriptstyle{P}}}\left(Q,R,S\right)\label{eq:WeightSelection}
	\end{equation*}characterizes the best LQR gain for tracking the desired STS movement from the pool of candidates, with respect to the performance metric. The values found after 8.2 days of computation with a workstation of 4 cores at $ 2.7 $[GHz] running Matlab Parallel Toolbox, are\begin{align*}
	Q^{\star} &= \operatorname{diag} \left( \left[80, \, 95, \, 95, \, 68, \, 90, \, 83 \right] \right),\\
	R^{\star} &= \operatorname{diag} \left( \left[1.0 \times 10^{-3}, \, 2.0 \times 10^{-4}, \, 6.0 \times 10^{-4}, \, 4.4 \times 10^{-3} \right] \right),\\
	S^{\star} &= \operatorname{diag} \left( \left[30, \, 37, \, 19, \, 29, \, 92, \, 82 \right] \right).
	\end{align*}Their matrix gain $ K_{\scriptscriptstyle{\mathrm{LQR}}}^{\star}\left(t\right) $ leads to $ J_{\mathrm{\scriptscriptstyle{P}}}^{\star} = 1.31 $. The significant improvement in tracking the input reference using this controller over the baseline is illustrated in Figure \ref{fig:FyComparison}, which exhibits the vertical force at the shoulders for simulations of the ascension phase for a set $ \mathcal{P}_{s} \subseteq \left[ \underline{p}, \overline{p} \right] $ of 500 parameters from a Latin Hypercube sampling (note that this set is different from $ \mathcal{P}_{b} $). When $ K_{\scriptscriptstyle{\mathrm{LQR}}}^{\star}\left(t\right) $ controls the system \eqref{eq:Nonlinear} (green lines) the deviations of the trajectories from the reference (red dashed line) are smaller than the ones achieved under the baseline controller (blue lines). Although this behavior is only expected at the time instants in $ T_{\mathrm{\scriptscriptstyle{P}}} $, it gracefully holds along the entire horizon. Similar improvement is registered for tracking $ \hat{x}(t) $, and $ \hat{y}(t) $.
	
	\section{Iterative Learning Control Algorithm \\ as a Proxy for User Action} \label{sec:ILC}
	
	Although the feedback controller \begin{equation*}
	u(t)=\hat{u}(t)-K_{\scriptscriptstyle{\mathrm{LQR}}}^\star(t)(x(t)-\hat{x}(t))
	\end{equation*}obtained in Section~\ref{subsec:performance metric} is the optimal choice from a batch of candidates, actual execution of the STS movement relies on the interaction of two agents driving different control inputs of the system. More precisely, of the $ 4 $-dimensional control input $ u=[\tau_h;\tau_s;F_x;F_y] $, only the torque at the hips $ \tau_h $ is executed by the actuators of the PLLO under the authority of its onboard computer, while the three other controls (torque $\tau_s$, horizontal $F_x$ and vertical $F_y$ forces at the shoulders) are to be applied solely by the user interacting with the ground through crutches. Thus, unlike the accurate computer implementation at the hips, the human implementation at the shoulders will rely on a limited perception of the state of the system due to paraplegia, and no preconceived knowledge of reference trajectories. Therefore, a controller which is optimal in simulations for \eqref{eq:Nonlinear}, when assuming perfect state feedback and actuation by the user, is not guaranteed to work experimentally. 
	
	The purpose of this section is thus to assess whether a proxy for the user actions can learn to cooperate with the LQR controller designed in Section~\ref{subsec:performance metric} through repeated trials and achieve a safe STS movement. For this, we choose to represent the interaction of the user with the PLLO using an Iterative Learning Control (ILC) algorithm.
	We believe that evaluating the performance of a controller for the hips combined with an ILC controller for the shoulders is a reasonable test prior to actual implementation of the PLLO.
	
	\subsection{Iterative Learning Control Algorithm}
	
	We start by rewriting the dynamics in \eqref{eq:ThreeLinkRobotModel} to better encompass the separate actions of the controller of the PLLO and the user. This is done by plugging the closed-loop expression for the input at the hips obtained from the finite horizon LQR controller in Section~\ref{subsec:LQR Design}, and leaving the input of the user in open-loop. For a more realistic representation, this model also incorporates saturation of the inputs; so far, constraints on their values have only been taken into account while solving the control allocation in~\eqref{eq:Allocation} for $ \hat{u} (t) $. We assume the user has healthy enough vestibular and visual systems, and adequate proprioception of the upper body to know the angular position $\theta_3(t)$ and velocity $\dot{\theta}_3(t)$ of the torso; and that the PLLO is instrumented to display in a monitor the real time trajectories of the position $(x_{\scriptscriptstyle{\mathrm{CoM}}}(t), y_{\scriptscriptstyle{\mathrm{CoM}}}(t))$ and velocity $(\dot{x}_{\scriptscriptstyle{\mathrm{CoM}}}(t), \dot{y}_{\scriptscriptstyle{\mathrm{CoM}}}(t))$ of the CoM, together with their references. This is in similar fashion as for the Robot Suit HAL, where users can see plots of the desired center of pressure and its actual position during training of the STS movement, in order to achieve proper synchronization with the device~\cite{Tsukahara2010}.
	
	Let $ D_{1} := [1 \; 0 \; 0 \; 0] $, and define the saturation of $ c \in \mathbb{R}^n $ over the interval $ [a,\,b] \subseteq \mathbb{R}^n $ as the element-wise $ \min / \max $ operation: \[ \operatorname{sat}\left( c, \, [a, \, b] \right) := \min \left( b, \, \max \left( a, \, c \right) \right).\] The input applied at the hips with state feedback from the finite horizon LQR controller, and within the limits of operation of the PLLO actuators, is \begin{align}
	\tau^{\star}_{\scriptscriptstyle{\mathrm{h}}} \left(t, x\right)&:= \operatorname{sat} \left( D_{1} \left( \hat{u}\left(t\right)-K^{\star}_{\scriptscriptstyle{\mathrm{LQR}}} \left( t \right) \left( x - \hat{x} \left(t\right) \right) \right) , \left[ D_{1} \underline{u}, \, D_{1} \overline{u} \right] \right). \label{eq:TauHipLQR}
	\end{align}
	
	Denote the loads at the shoulders as $ \mu := [\tau_{s}; F_{x}; F_{y}] \in \mathbb{R}^3 $, and the output measured by the user as $ \Upsilon := [ \theta_3; \; x_{\scriptscriptstyle{\mathrm{CoM}}}; \; y_{\scriptscriptstyle{\mathrm{CoM}}}; \; \dot{\theta}_3; \; \dot{x}_{\scriptscriptstyle{\mathrm{CoM}}}; \; \dot{y}_{\scriptscriptstyle{\mathrm{CoM}}}] \in \mathbb{R}^6 $. Taking $ D_{2} \in \mathbb{R}^{3 \times 4}$ as $ D_{2}:= \left[0_{3 \times 1} \; I_3 \right] $, where $ I_3 $ is the identity matrix, we define $ \hat{\mu}(t) := D_{2} \hat{u}(t) $, $ \underline{\mu} := D_{2} \underline{u} $, and $ \overline{\mu} := D_{2} \overline{u} $. $ \Upsilon $ can be determined from the state $ x $ and parameter $ p $ with a mapping denoted as $ \Psi: \mathbb{R}^{6} \times \mathbb{R}^{12} \rightarrow \mathbb{R}^{6} $ using the kinematic equations of the CoM of the three-link robot in Appendix \ref{appendix:CoMKinematics}. Plugging $ \tau^{\star}_{\scriptscriptstyle{\mathrm{h}}} \left(t, x\right) $ into \eqref{eq:ThreeLinkRobotModel}, the nonlinear dynamics of the system with user input $ \mu $ and output $ \Upsilon $ are
	\begin{align}
	\dot{x}\left(t\right) & =f\left(x\left(t\right),p,\left[ \begin{array}{c}
	\tau^{\star}_{h} \left(t, x\right) \\
	\mu\left( t \right)
	\end{array} \right]\right) =: \Xi \left(t, x, p, \mu \right) \nonumber \\
	\Upsilon \left(t\right) &= \Psi \left(x(t), p \right). \label{eq:ThreeLinkILC}
	\end{align}The desired trajectory for $ \Upsilon \left(t\right) $ during the ascension phase of the STS movement is computed as $ \hat{\Upsilon}\left(t\right) = \Psi \left(\hat{x}\left(t\right), \hat{p} \right) $.
	
\begin{figure}
		\begin{centering}
			\includegraphics[width=12cm]{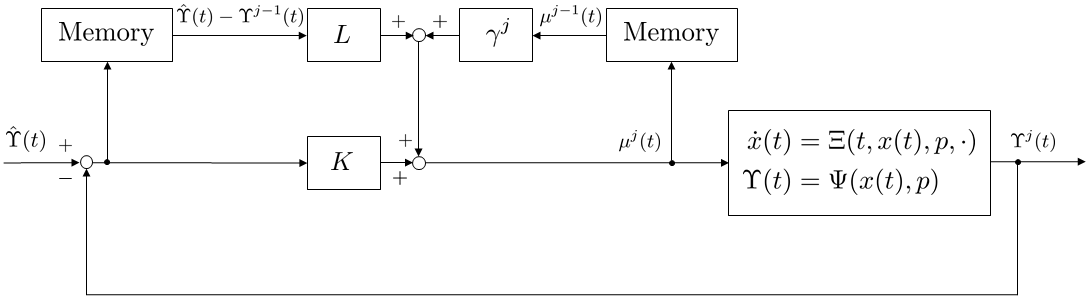}
			\par\end{centering}
		\caption{Basic ILC algorithm with feedback gain $ K $, feedforward gain $ L $, and recalling matrix $ \gamma^j $.\label{fig:ILC}}
	\end{figure}

	The algorithm to emulate the loads applied at the shoulders by a user, over $ N $ ascension attempts with the PLLO, is built upon the general current-iteration ILC referred to in \cite{Bristow2006}. Translating such control strategy to our problem, for successive iterations indexed by $ j \in \{ 1, \, \ldots, \, N \} $, the user input $ \mu^j(t) $ at $ t \in [t_0,\,t_f] $ is given by\begin{align}
	\mu^{j}(t) &:= \gamma_j \mu^{j-1}(t) + L \left( \hat{\Upsilon}(t)-\Upsilon^{j-1}(t) \right) + K \left( \hat{\Upsilon}(t)-\Upsilon^{j}(t) \right), \label{eq:BasicILC}
	\end{align}where $ \gamma_j \in \mathbb{R}^{3 \times 3} $, $ L,K \in \mathbb{R}^{ 3 \times 6 } $, and $ \Upsilon^{0}(t) := \hat{\Upsilon}(t) $. The block diagram in Figure~\ref{fig:ILC} shows that the feedforward component of this basic learning algorithm consists of two terms relying on the memory of the user about the past-iteration $ j-1 $. The feedforward gain $ L $ modifies the input $ \mu^{j}(t) $ according to the error of the output $ \hat{\Upsilon}(t) - \Upsilon^{j-1}(t) $, while the recalling matrix $ \gamma_j $ is inspired by \cite{Arif1999} and is used to capture the ability of the user to remember and execute $ \mu^{j-1}(t) $. If $ \gamma_j = I_3 $ then $ \mu^{j-1}(t) $ is perfectly incorporated into $ \mu^{j}(t) $, but if $ \gamma_j \neq I_3 $ we will interpret the mismatch between the values of $\gamma_j \mu^{j-1}(t)$ and $\mu^{j-1}(t)$ to be the consequence of either a memory flaw, an imperfect execution of the required loads at the shoulders, or a combination of both. The feedback component changes the input of the user by multiplying the error of the outputs at the current-iteration $ j $ by the feedback gain $ K $. Although at the beginning of training there is no preconceived notion of the input that needs to be exerted to attempt a STS movement with the PLLO, we consider that the values $ \hat{\mu}(t_0) $ and $ \hat{\mu}(t_f) $ are known to the user, so that the ILC can be initialized with the linear interpolation\begin{align}
	\mu^{0}(t) &= \frac{\hat{\mu}(t_f)-\hat{\mu}(t_0)}{t_f-t_0} (t-t_0) + \hat{\mu}(t_0). \label{eq:mu0}
	\end{align}
	
	When the input $ \mu^{j}(t) $ from the basic ILC algorithm in \eqref{eq:BasicILC} acts on \eqref{eq:ThreeLinkILC}, the state trajectory $ x(t) $ might lead to a configuration of the links of the PLLO which is harmful for the user, or even mechanically impossible to reach. To elaborate on this situation, we delimit the feasible ranges of motion for the user and the PLLO with the state bounds\begin{align*}
	\underline{x} &:= \left[ 80 \, \left[ ^{\circ} \right]; \; -120 \, \left[ ^{\circ} \right]; \; 0 \, \left[ ^{\circ} \right]; \; -20 \, \left[ ^{\circ}/ \mathrm{s} \right]; \; -5 \, \left[ ^{\circ}/ \mathrm{s} \right]; \; -70 \, \left[ ^{\circ}/ \mathrm{s} \right] \right], \\
	\overline{x} &:= \left[ 120 \, \left[ ^{\circ} \right]; \; 0 \, \left[ ^{\circ} \right]; \; 130 \, \left[ ^{\circ} \right]; \; 10 \, \left[ ^{\circ}/ \mathrm{s} \right]; \; 60 \, \left[ ^{\circ}/ \mathrm{s} \right]; \; 20 \, \left[ ^{\circ}/ \mathrm{s} \right] \right];
	\end{align*}and let $ t_{s}^{j} \in [t_0,\,t_f] $ be the maximum value such that the state at iteration $j$ satisfies $ x(t) \in \left[ \underline{x}, \, \overline{x} \right] $ for all $ t \in [t_0,\, t_{s}^{j} ] $. To account for situations where the user would need to abort execution of the STS movement due to safety concerns, we stop the ongoing iteration when $ x (t) $ goes out of bounds, reset the state of the system to the initial condition $ x_0 $, and proceed to the next one.
	
	Suppose the human input caused the STS movement in the past-iteration $ j-1 $ to stop prematurely, so that $ t_0 < t_{s}^{j-1} < t_f $. In this scenario, the error $ \hat{\Upsilon}(t)-\Upsilon^{j-1}(t) $ and $ \mu^{j-1}(t) $ in \eqref{eq:BasicILC} only exist for $ t \in \left[ t_0, t_{s}^{j-1} \right] $. In order to compute $ \mu^{j}(t) $ for $ t > t_{s}^{j-1} $, define $ \Gamma^{j}(t) \in \mathbb{R}^3 $ for $ j \in \{ 1, \, \ldots, \, N \} $ as \begin{align}
	\Gamma^{j}(t) &:= \begin{cases}
	\gamma_j \mu^{j-1}(t) + L \left( \hat{\Upsilon}(t)-\Upsilon^{j-1}(t) \right) + K \left( \hat{\Upsilon}(t)-\Upsilon^{j}(t) \right)  , & t < t_{s}^{j-1} \\
	\alpha^{j} t + \beta^{j}, & t \geq t_{s}^{j-1}
	\end{cases}, \label{eq:TrainingILCGamma}
	\end{align}where $ \alpha^{j}:= \frac{\hat{\mu}(t_f)-\mu^{j-1}(t^{j-1}_s)}{t_f-t^{j-1}_s} $ and $ \beta^{j}:= \hat{\mu}(t_f) - \alpha^{j} t_f $ implement a linear extrapolation of time for the human input, between $ \mu^{j-1}(t^{j-1}_s) $ and $ \hat{\mu}(t_f) $. This is to simulate the torque and forces that the user would begin to apply at $ t^{j-1}_s $ while attempting to counteract the negative effects of the past-iteration input on the ascension. For initialization purposes, $ t_{s}^{0}:= t_f $. 
	
	Adding saturation to the extremal loads that the user is physically capable of exerting at the shoulders, we complete our ILC algorithm as:\begin{align}
	\mu^{j}(t) &:= \operatorname{sat} \left( \Gamma^{j}(t), \, \left[ \underline{\mu}, \, \overline{\mu} \right] \right). \label{eq:TrainingILCmu}
	\end{align}
	
	\subsection{Tuning of the ILC gains}
	\label{sub:DFO}
	We reason that if we can find ILC gains for simulating a realistic ascension phase after a limited number of iterations, then a real user, who has a more complex learning process, would be able to coordinate with the controller for the hips of the PLLO through training and complete a successful STS movement.
	
	For the safety of the users, we should refrain from using any experimental setting that involves exposing them to potentially harmful actions of unproven controllers. Therefore, instead of running identification techniques, we resort to a reinforcement learning approach to numerically search for the values of the feedforward $ L $ and feedback $ K $ gains in \eqref{eq:TrainingILCGamma}.
	
	Define the $ j^{\mathrm{th}} $ iteration cost\begin{align}
	J^{j}_{\scriptscriptstyle{\mathrm{L}}} &:= \begin{cases}
	\infty, & t_{s}^{j} < t_f\\
	\int_{t_0}^{t_f} \left( \left\Vert \hat{\Upsilon}(t)-\Upsilon^{j}(t)\right\Vert_{2} +  w_{\mu} \left\Vert \dot{\mu}^{j}(t) \right\Vert_{2} \right) \, dt, & t_{s}^{j} = t_f 
	\end{cases}, \label{eq:JL}
	\end{align}where the weight $ w_{\mu}:= 10^{-4} $ is used to account for the different units of $ \Upsilon^{j}(t) $ and $ \dot{\mu}^{j}(t) $. If a particular choice of gains causes the final iteration $N$ to stop the STS movement prematurely ($ t_{s}^{N} < t_f $), it must be discarded for modeling the behavior of the user. Otherwise (if $t_{s}^{N} = t_f$), the quality of its corresponding ILC algorithm should be assessed, based on the deviation of the output in \eqref{eq:ThreeLinkILC} from its desired trajectory $ \hat{\Upsilon}(t) $, and the rate of change of the input $ \dot{\mu}^{N}(t) \in \mathbb{R}^3 $. With the nominal parameter $ p = \hat{p} $ in \eqref{eq:ThreeLinkILC}, a time step of $ 4 \, [\mathrm{ms}] $ for computing a discrete version of the iteration cost $ J^{j}_{\scriptscriptstyle{\mathrm{L}}} $, $ \gamma_j = I_3 $ for all $ j \in \{ 1, \, \ldots, \, N \} $, and $ N := 30 $ iterations, we select the gains in \eqref{eq:TrainingILCGamma} as:\begin{align}
	K^{\star},\,L^{\star} = \underset {K,L \in \mathbb{R}^{3 \times 6}}{\arg\min}J^{N}_{\mathrm{\scriptscriptstyle{L}}}\left(K,L\right).\label{eq:DFO}
	\end{align}
	
	There are two major difficulties for computing the solution to \eqref{eq:DFO}. The first is that the problem is in general non-convex and hence global minimization is intractable. The second is that computing derivatives with respect to $K$ and $L$ is cumbersome. The first issue is typically dealt with in practice via heuristic local search; we find this to be effective for our problem. To deal with the second issue, we apply a standard technique from the optimization literature for minimizing a function using only black-box function calls, which we describe briefly.
	
	Suppose we want to minimize $g : \mathbb{R}^n \rightarrow \mathbb{R}$ over $\mathbb{R}^n$. For $\sigma > 0$, we define a smoothed function $G_\sigma(\eta) := \mathbb{E}_{\xi}[ g(\eta + \sigma \xi) ]$, where $\xi \sim \mathcal{N}(0, I_n)$ is an isotropic Gaussian random vector, and $\mathbb{E}$ denotes the expectation. Under reasonable regularity conditions on $g$, a standard calculation shows that the gradient of $G_\sigma$ is given by\begin{align*}
	\nabla G_\sigma(\eta) = \mathbb{E}_{\xi}\left[ \frac{g(\eta + \sigma \xi) - g(\eta - \sigma \xi)}{2\sigma} \xi \right] \:.
	\end{align*}
	That is, we can differentiate $G_\sigma(\eta)$ by only using function calls of $g$. We can interpret this as a finite-difference method applied in a random direction. Furthermore, it is clear that as $\sigma \to 0$, $G_\sigma(\eta)$ approaches $g(\eta)$. Hence, optimizing $G_\sigma$ is a reasonable proxy for optimizing $g$; this is made formal in \cite{nesterov}. 
	The most basic way to apply derivative free
	optimization is to run stochastic gradient descent:
	\begin{align}
	\eta_{k+1} = \eta_k - \rho_k \frac{g(\eta_k + \sigma \xi_k) - g(\eta_k - \sigma \xi_k)}{2\sigma} \xi_k \:, \label{eq:StocGradDesc}
	\end{align}where $\{ \rho_k \}_{k \geq 0}$ is an appropriate sequence of step sizes and $\{\xi_k\}_{k \geq 0}$ is an independent and identically distributed sequence of $\mathcal{N}(0, I_n)$ random vectors.
	We apply a slightly modified version of
	\eqref{eq:StocGradDesc} as described in~\cite{mania18}. First, at every 
	iteration $k$ we draw $B$ random directions $\{ \xi_k^{i} \}_{i=1}^{B}$. We then sort the indices $i=1, ..., B$ in ascending order with the value assigned to each index given by $\min \left( g(\eta_k + \sigma \xi_k^i), g(\eta_k - \sigma \xi_k^i) \right) $, and compute the update direction as:
	\begin{align}
	\eta_{k+1} = \eta_k - \frac{\rho_k}{B_t} \sum_{i=1}^{B_t} \frac{g(\eta_k + \sigma \xi_k^{(i)}) - g(\eta_k - \sigma \xi_k^{(i)})}{2\sigma_{B_t}} \xi_k^{(i)} \:.
	\end{align}
	Here, $B_t \leq B$, $\xi_k^{(i)}$ denotes the sorted directions, and $\sigma_{B_t}$ denotes the \emph{empirical} standard deviation of the $2B_t$ costs used in the update. 
	
	We run this method for $10,000$ iterations with $g = J^{N}_{\mathrm{L}}$, $B=30$, $B_t=10$, $\sigma=0.01$, and $\rho_k \equiv 0.04$ on a machine with 72 physical cores running at 2.10 [GHz]. 
	After approximately two days of computation, we obtained the gains
	\begin{align*}
	K^{\star} = &\left[\begin{array}{cccccc}
	-100.6 & -53.26 & 71.59 & -123.9 & -208.5 & 66.02\\
	57.98 &  -21.06 & -47.92 & -24.67 & 166.9 & -31.92\\ 
	28.86 & 20.20 & 139.9 & 46.62 & -29.75 & 123.5
	\\
	\end{array} \right],
	\end{align*}
	\begin{align*}
	L^{\star} = &\left[ \begin{array}{cccccc}
	-0.6381 & 2.376 & 46.31 & -2.514 & -5.255 & -28.89 \\
	-2.948 & -29.34 & -25.24 & 2.851 & -3.559 & 24.48\\ 
	-73.09 & -58.25 & 114.1 & 19.82 & 13.09 & 126.9
	\end{array}\right],
	\end{align*}with a cost of $J^{\star}_{\scriptscriptstyle{L}}:=8.66$. As a reference, plugging $ \Upsilon^{N}(t) = \hat{\Upsilon}(t) $ and $ \dot{\mu}^{N}(t) = \dot{\hat{\mu}}(t) $ into \eqref{eq:JL} gives $\hat{J}_{\scriptscriptstyle{L}} := 8.75$. 
	Figures \ref{fig:Theta1}-\ref{fig:Fy} show in black the phase planes for the state $x$, position and velocity of the CoM in the sagittal plane, and input trajectories $ u(t) $ of the three-link robot model in \eqref{eq:ThreeLinkILC} simulated under $\mu^{30}(t)$.
	
	The ILC algorithm with gains $ K^{\star} $ and $ L^{\star} $ achieves almost perfect tracking of $ \hat{\Upsilon}(t) $; as a consequence, Figures \ref{fig:Theta3}, \ref{fig:PosCoM}, and \ref{fig:VelCoM} show the trajectories in black essentially overlapping with it. Deviations from the reference in the phase planes $ \theta_1-\dot{\theta}_1 $ (Figure \ref{fig:Theta1}) and $ \theta_2-\dot{\theta}_2 $ (Figure \ref{fig:Theta2}) are expected, since they are not penalized in \eqref{eq:JL}. Nevertheless, $\theta_1$ is off the vertical at the end of the ascension just by $0.5[^{\circ}]$, with the absolute values of both angular velocities less than $1.2[^\circ/s]$, which should not compromise the ability of a controller for the stabilization phase to reach the standing position with ease. As $\theta_2(t)$ remains less than zero for the entire trajectory, there is no hyperextension of the knees, and thus the input $\mu^{30}(t)$ should not pose a threat to the physical integrity of the user. 
	Even though the tracking errors for the angular position and velocity of the shank and thigh links do not directly affect the computation of $\mu^{j}(t)$ in \eqref{eq:TrainingILCmu}, they do determine (together with the tracking errors of the angular position and velocity of the torso) the value of the torque at the hips through the state feedback of the LQR, hence causing it to differ from $\hat{\tau}_{h}(t)$ in Figure \ref{fig:TauHip}. It is especially interesting that although we did not consider $\hat{\mu}(t)$ in \eqref{eq:JL}, both the torque and horizontal force at the shoulders in Figures \ref{fig:TauShoulder} and \ref{fig:Fx} follow their reference trajectories reasonably well, within errors of 15[N.m], and 10[N], respectively. The absolute value of the torque applied at the hips in Figure \ref{fig:TauHip} is in general greater than $\hat{\tau}_{h}(t)$, which compensates for lower vertical forces attained by $ \mu^{30}_3(t)$ in Figure \ref{fig:Fy} relative to $ \hat{F}_{y}(t) $. From the rate of change $ \dot{\mu}^{30}_3(t)$ observed in Figure \ref{fig:Fy}, we can infer that $J^{\star}_{\scriptscriptstyle{L}} < \hat{J}_{\scriptscriptstyle{L}}$ is mostly due to the difference of its values over time with respect to $\dot{\hat{F}}_y(t)$. Although $\hat{F}_y(t)$ remains constant for about 1[s], it decreases 165[N] in 1.4[s], while $ \mu^{30}_3(t)$ decreases 155[N] over 2.5[s] for a lower average rate of change.
	
	\subsection{Robustness Evaluations}
	
	The results discussed above indicate that the ILC in \eqref{eq:TrainingILCmu} is able to successfully coordinate with the LQR controller driving the actuators at the hips in \eqref{eq:TauHipLQR} to complete the desired ascension phase with no risk of sit-back or step failures \cite{Eby2006}. Moreover, it does so exhibiting input trajectories that could be realistically executed by both the PLLO and the user after 30 learning iterations. Since the gains $K^{\star}$ and $L^{\star}$ that make this behavior possible were found considering a constant recalling matrix $\gamma_j=I_3$ in \eqref{eq:TrainingILCGamma} across every iteration, and the nominal value of the parameter $\hat{p}$ in \eqref{eq:ThreeLinkILC}, the purpose of this section is to evaluate the robustness of the ILC algorithm to perform the STS movement in two scenarios: imperfect recalling and execution of $\mu^{j-1}(t)$, and parameter uncertainty.
	
\begin{figure}[H]
		\centering
		\subfloat[Phase plane of trajectories $ \theta_1(t) $ and $ \dot{\theta}_1(t) $. \label{fig:Theta1}]{\includegraphics[width=8cm]{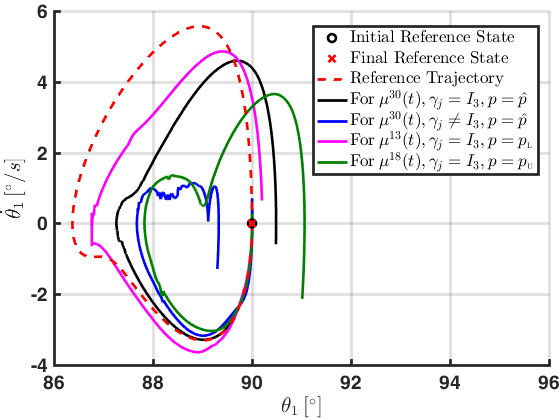}} \hspace{1.5cm}
		\subfloat[Phase plane of trajectories $ \theta_2(t) $ and $ \dot{\theta}_2(t) $. \label{fig:Theta2} ]{\includegraphics[width=8cm]{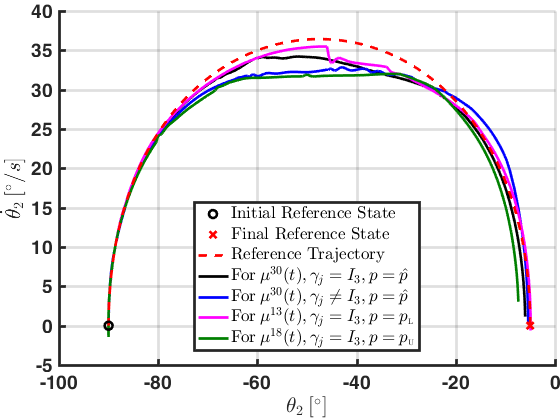}}
		\\
		\subfloat[Phase plane of trajectories $ \theta_3(t) $ and $ \dot{\theta}_3(t) $. \label{fig:Theta3}]{ \includegraphics[width=8cm]{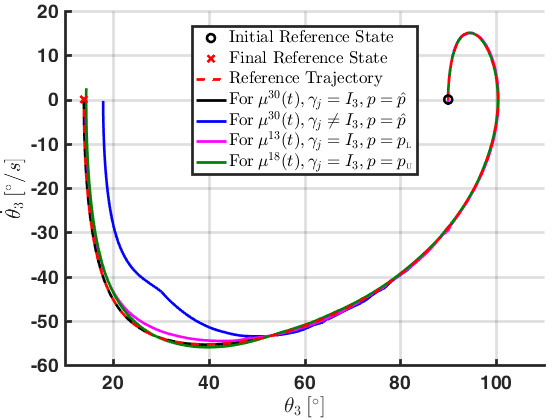}} \hspace{1.5cm}
		\subfloat[Trajectories for the position of the CoM in the sagittal plane.\label{fig:PosCoM}]{ \includegraphics[width=8cm]{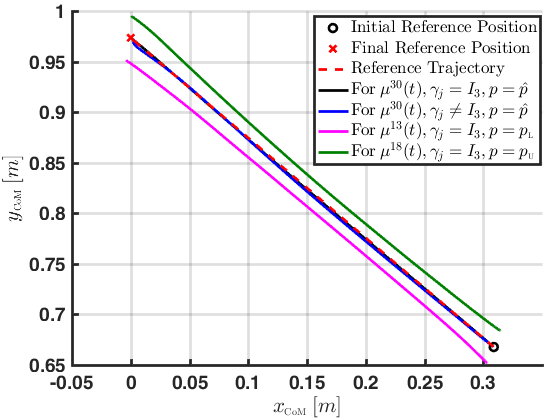}}
		\\
		\subfloat[Trajectories for the velocity of the CoM in the sagittal plane. \label{fig:VelCoM}]{ \includegraphics[width=8cm]{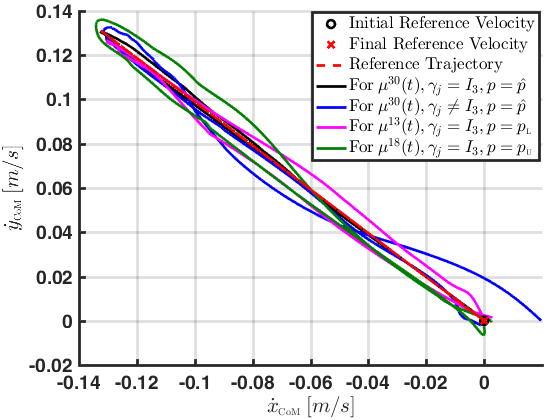}} \hspace{1.5cm}
		\subfloat[Torque applied at the hips by the LQR controller. \label{fig:TauHip}]{ \includegraphics[width=8cm]{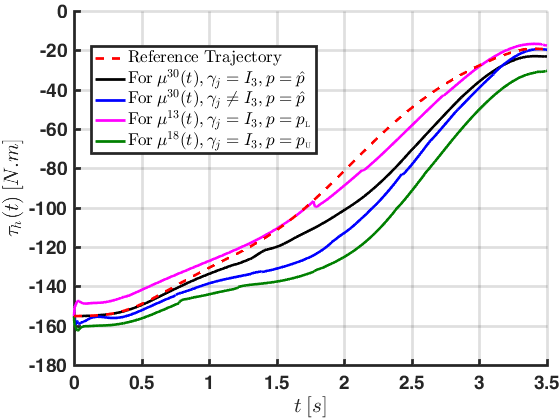}}
		\caption{Phase planes for the state $x$, position and velocity of the CoM in the sagittal plane, and torque applied at the hips by the LQR controller of the PLLO.}
	\end{figure}

\begin{figure}[H]
		\centering
		\subfloat[Torque applied at the shoulders by the ILC algorithm. \label{fig:TauShoulder}]{\includegraphics[width=8cm]{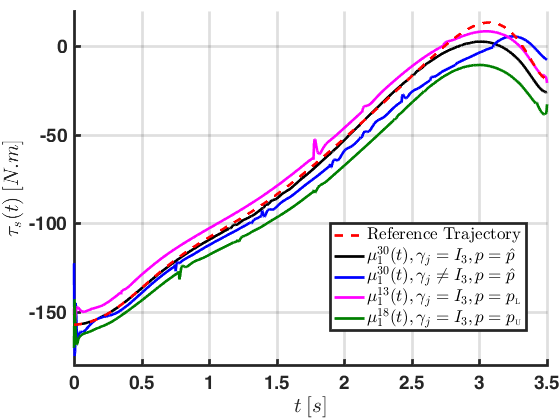}} \hspace{1.5cm}
		\subfloat[Horizontal force applied at the shoulders by the ILC algorithm. \label{fig:Fx}]{\includegraphics[width=8cm]{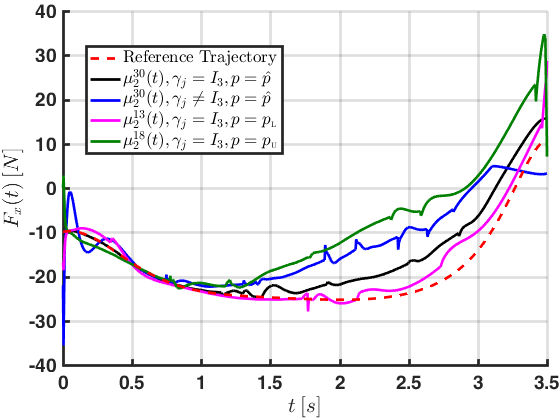}}
		\\
		\subfloat[Vertical force applied at the shoulders by the ILC algorithm. \label{fig:Fy}]{ \includegraphics[width=8cm]{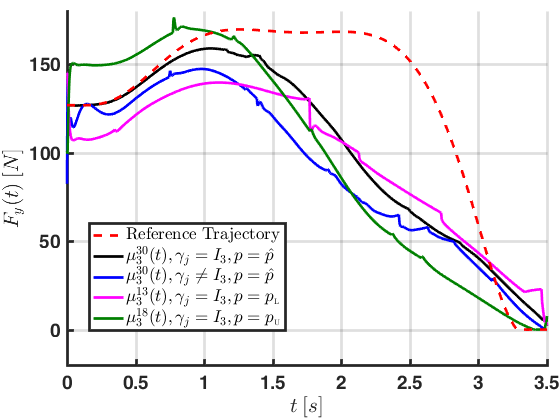}}
		\caption{Loads applied at the shoulders by the ILC algorithm.}
	\end{figure}

	For the imperfect recalling and execution of $\mu^{j-1}(t)$ we plug the iteration-varying matrix $\gamma_j = I_3+q^{j-1}\vartheta_{j} $ in \eqref{eq:BasicILC}, with $q:=0.8$, and randomly sample the entries of $ \vartheta_j \in \mathbb{R}^{3 \times 3} $ at every iteration $j \in {1,\ldots,N}$ within the interval $ \left[ -0.05, 0.05 \right] $. While the off-diagonal entries in $\vartheta_{j}$ couple the loads at the shoulders, the decay to zero of the power function $ q^{j-1} $ as the number of trials increases, captures the idea that a user would eventually recall and apply the appropriate values of $ \mu^{j-1}(t) $. With the nominal parameter value $ \hat{p} $, starting the learning algorithm from the linear interpolation in \eqref{eq:mu0}, and applying $30$ iterations, we obtained the behavior of the system \eqref{eq:ThreeLinkILC} shown in blue in Figures \ref{fig:Theta1}-\ref{fig:Fy}. The degraded tracking of $ \hat{\Upsilon}(t) $ and $\hat{x}(t)$ is evident in Figures \ref{fig:Theta1}-\ref{fig:VelCoM}. However, the trajectory of the CoM position in Figure \ref{fig:PosCoM} shows that it is still possible to complete the ascension. Furthermore, Figure \ref{fig:Theta2} affirms that the integrity of the knee joints will be preserved. Judging from the deviation from the reference in the phase plane for the angular position and velocity of the shanks in Figure \ref{fig:Theta1}, and most importantly the non-zero final velocity of the CoM observed in Figure \ref{fig:VelCoM}, we predict that the stabilization phase of this STS movement would be more challenging than the one obtained for $ \mu^{30}(t)$ in the previous section, when $ \gamma_j = I_3 $. The control inputs of the PLLO (Figure \ref{fig:TauHip}) and those obtained in the ILC algorithm (Figures \ref{fig:TauShoulder}-\ref{fig:Fy}) remain between the bounds $[\underline{u},\overline{u}]$, verifying that $K^{\star}$ and $L^{\star}$ still attain realistic trajectories after the same number of iterations with $ \gamma_j \neq I_3 $, although their oscillations and sudden changes do lead to an increased value for the cost in \eqref{eq:JL} of $J^{30}_L=40$. 
	
	We now study the effect of parameter uncertainty after the ILC algorithm has completed $30$ iterations under the nominal value of the parameter $\hat{p}$, and constant recalling matrix $\gamma_{j} = I_3$. For this purpose, we set the new $\mu^{0}(t)$ in \eqref{eq:TrainingILCGamma} equal to the trajectories in black from Figures \ref{fig:TauShoulder}-\ref{fig:Fy}, and simulate the system in \eqref{eq:ThreeLinkILC} under the action of \eqref{eq:TrainingILCmu} for two different parameter values:\begin{align*}
	p_{\mathrm{\scriptscriptstyle{L}}} := [& 9.2\left[ \mathrm{kg} \right]; \; 11.2\left[ \mathrm{kg} \right];\; 42.3\left[ \mathrm{kg} \right]; 1.10\left[ \mathrm{kg \cdot m^2} \right];\; 0.49\left[ \mathrm{kg \cdot m^2} \right];\; 2.40\left[ \mathrm{kg \cdot m^2} \right]; \ldots \nonumber \\ 
	& 0.529\left[ \mathrm{m} \right]; \; 0.409\left[ \mathrm{m} \right]; \; 0.519\left[ \mathrm{m} \right]; 0.23 \left[ \mathrm{m} \right]; \; 0.17\left[ \mathrm{m} \right]; \; 0.24\left[ \mathrm{m} \right]],
	\end{align*}\begin{align*}
	p_{\mathrm{\scriptscriptstyle{U}}} := [& 10.2\left[ \mathrm{kg} \right]; \; 13.2\left[ \mathrm{kg} \right];\; 46.8\left[ \mathrm{kg} \right]; 1.21\left[ \mathrm{kg \cdot m^2} \right];\; 0.54\left[ \mathrm{kg \cdot m^2} \right];\; 2.65\left[ \mathrm{kg \cdot m^2} \right]; \ldots \nonumber \\ 
	& 0.531\left[ \mathrm{m} \right]; \; 0.411\left[ \mathrm{m} \right]; \; 0.521\left[ \mathrm{m} \right]; 0.30 \left[ \mathrm{m} \right]; \; 0.23\left[ \mathrm{m} \right]; \; 0.28\left[ \mathrm{m} \right]].
	\end{align*}All entries match the bounds in Table \ref{tab:UncertainP}, except for the ones representing the lengths of the links, which come from subtracting (for $p_{\mathrm{\scriptscriptstyle{L}}}$), and adding (for $p_{\mathrm{\scriptscriptstyle{U}}}$) 1[mm] to the nominal lengths in $\hat{p}$. This choice puts more emphasis on studying changes in performance stemming from fluctuations of the total mass of the user, rather than from variations on the length of the links, since, in practice, the latter are only expected to occur due to wear of mechanical components after an extended period of use, while the former are bound to happen several times through the day. Keeping track of the cost $J^{j}_{\scriptscriptstyle{\mathrm{L}}}$ over $j=\{1,\ldots,30\}$, we identify the iterations where the ILC algorithm attains the minimum values, and present their corresponding simulations in Figures \ref{fig:Theta1}-\ref{fig:Fy}. The results for $p_{\mathrm{\scriptscriptstyle{L}}}$ after $13$ iterations are in magenta, while the ones for $ p_{\mathrm{\scriptscriptstyle{U}}}$ after $18$ are in green. The associated costs are $J^{13}_{\scriptscriptstyle{L}}:=27.9$ and $J^{18}_{\scriptscriptstyle{L}}:=29.1$.
	
	Since $\hat{x}(t)$ and $\hat{\Upsilon}(t)$ are determined based on the nominal value $\hat{p}$, the reference for the position of the CoM in Figure \ref{fig:PosCoM} cannot be perfectly tracked with the mismatch in the parameter values. Their simulations run approximately parallel to the reference, with the lower bounds for $l_{c_{i}}$ in $p_{\mathrm{\scriptscriptstyle{L}}}$ causing its trajectory to be below it, and the upper bounds for $l_{c_{i}}$ in $p_{\mathrm{\scriptscriptstyle{U}}}$ putting its trajectory above. Although the velocities of the CoM for both parameter values in Figure \ref{fig:VelCoM} also deviate from the reference, zero velocity is achieved at the end of the movements, which together with the behavior observed in the phase plane in Figure \ref{fig:Theta2}, proves that the proxy for the user action can safely complete the ascension phase. According to the larger offsets from the final desired state that exhibit the trajectories for $p_{\mathrm{\scriptscriptstyle{U}}}$ with respect to $p_{\mathrm{\scriptscriptstyle{L}}}$ in Figures \ref{fig:Theta1}-\ref{fig:Theta3}, we predict that the stabilization phase for a situation where the total mass of the user increases by $5\%$ of its nominal value, might be more challenging than the one where the total mass decreases by the same amount. Figures \ref{fig:TauHip} and \ref{fig:TauShoulder} show that an increased mass of the wearer also requires larger contributions from the torque executed by the PLLO at the hips and the torque applied by the user at the shoulders. The oscillations and sudden changes of the force profiles (in magenta and green) in Figures \ref{fig:Fx} and \ref{fig:Fy}, together with the deviations from $ \hat{\Upsilon}(t) $ in Figures \ref{fig:Theta3}-\ref{fig:VelCoM}, contribute to the larger values of $J^{13}_{\scriptscriptstyle{L}}$ and $J^{18}_{\scriptscriptstyle{L}}$ relative to $J^{\star}_{\scriptscriptstyle{L}} $. 
	It is interesting that the most abrupt changes of $ \dot{\mu}^{13}(t) $ happen when the velocity of the CoM achieves its maximum (approximately at 1.77[s]), and in the last 0.1[s] of the ascension; where $ \dot{\mu}^{18}(t) $ also experiences its most significant changes.   
	To further analyze the effect of parameter uncertainty on the performance of our proposed ILC algorithm in \eqref{eq:TrainingILCmu}, we also considered 500 random parameter values $p \in [p_{\scriptscriptstyle{L}},p_{\scriptscriptstyle{U}}]$ from a Latin Hypercube sampling to perform analogous evaluations to the ones for $p_{\mathrm{\scriptscriptstyle{L}}}$ and $p_{\mathrm{\scriptscriptstyle{U}}}$. All simulations exhibited safe, realistic, and successful ascending STS movements.
	
	\section{Conclusions}
	
	The paper presented a procedure to obtain reference trajectories for the ascension phase of STS movements for a minimally actuated PLLO with sagittal symmetry, to design a pool of finite horizon LQR controllers for tracking such trajectories, to choose the best controller relative to a robust performance metric in the presence of parameter uncertainty, and to evaluate through simulation if it would be adequate for implementation using an ILC algorithm as a surrogate for the human input. The gains of the ILC algorithm are tuned via reinforcement learning under nominal conditions, and perfect recalling/execution of the input trajectory from previous trials.  
	
	
	This procedure can be custom tailored to the actuators available in the PLLO, and the height and weight of the user. Nevertheless, since they rely completely on the three-link robot model, future work might include exploring control schemes that also take into account model uncertainties in addition to the parameter uncertainties already considered in this paper.
	
	Given the complexity of the closed-loop dynamics of the system, estimating the sensitivity bounds for computing the over-approximation functions in the performance metric required a time consuming sampling of the parameter interval. Improving this procedure is another appealing research direction.  
	
	Our most important conclusion is that the best controller with respect to the robust performance metric works in harmony with the ILC algorithm substituting for the shoulder actions. It is remarkable that we incur only limited performance degradation, even in the presence of extreme parameter uncertainties, and under scenarios of flawed memory and lack of coordination at the shoulders. 
	
	These favorable conclusions encourage further steps towards human testing. We believe that our procedure can set a good benchmark to systematically choose actuators of PLLOs to fit a larger variety of users, estimate the time and effort required for training and develop a protocol for better assessing the robustness of the STS movement in clinical trials. This would then help to close the gap between PLLOs and standing wheelchairs, which still remain the most reliable mobility solution for patients with complete paraplegia.
	\newpage
	\appendices
	%
	%
	\section{Terms in Euler-Lagrange Equations} \label{appendix:EulerLagrange}
	
	For notational convenience in the following sections, denote $ c_{i} :=\cos\theta_{i} $, $ c_{ij} :=\cos\left(\theta_{i}+\theta_{j}\right) $, $ c_{ijk} :=\cos\left(\theta_{i}+\theta_{j}+\theta_{k}\right) $, and similarly for $ \sin\left(\cdot\right) $. Also, define the coefficients\[\begin{array}{lll}
	k_{0}\left(p\right):=\left(m_{1}+m_{2}+m_{3}\right)^{-1}, &  & k_{1}\left(p\right):=l_{\scriptstyle{\mathrm{c}_1}}m_{1}+l_{1}m_{2}+l_{1}m_{3},\\
	k_{2}\left(p\right):=l_{\scriptstyle{\mathrm{c}_2}}m_{2}+l_{2}m_{3}, &  & k_{3}\left(p\right):=l_{\scriptstyle{\mathrm{c}_3}}m_{3},
	\end{array}\]whose explicit dependence with respect to the parameter $ p $ will be omitted onwards for compactness.
	
	The entries of the symmetric mass matrix $M\left(\theta,p\right)\in\mathbb{R}^{3\times3}$ in \eqref{eq:EulerLagrange} are\begin{align*}
	M_{11} =&  I_{1}+I_{2}+I_{3}+l_{\scriptstyle{\mathrm{c}_1}}^{2}m_{1}+m_{2}\left(l_{1}^{2}+2l_{1}l_{\scriptstyle{\mathrm{c}_2}}c_{2}+l_{\scriptstyle{\mathrm{c}_2}}^{2}\right)+ m_{3}\left(l_{1}^{2}+2l_{1}l_{2}c_{2}+2l_{1}l_{\scriptstyle{\mathrm{c}_3}}c_{23}+l_{2}^{2}+2l_{2}l_{\scriptstyle{\mathrm{c}_3}}c_{3}+l_{\scriptstyle{\mathrm{c}_3}}^{2}\right)\\
	M_{12} =& I_{2}+I_{3}+l_{\scriptstyle{\mathrm{c}_2}}m_{2}\left(l_{1}c_{2}+l_{\scriptstyle{\mathrm{c}_2}}\right) + m_{3}\left(l_{1}l_{2}c_{2}+l_{1}l_{\scriptstyle{\mathrm{c}_3}}c_{23}+l_{2}^{2}+2l_{2}l_{\scriptstyle{\mathrm{c}_3}}c_{3}+l_{\scriptstyle{\mathrm{c}_3}}^{2}\right)\\
	M_{13} =& I_{3}+l_{\scriptstyle{\mathrm{c}_3}}m_{3}\left(l_{1}c_{23}+l_{2}c_{3}+l_{\scriptstyle{\mathrm{c}_3}}\right)\\
	M_{22} =& I_{2}+I_{3}+l_{\scriptstyle{\mathrm{c}_2}}^{2}m_{2}+m_{3}\left(l_{2}^{2}+2l_{2}l_{\scriptstyle{\mathrm{c}_3}}c_{3}+l_{\scriptstyle{\mathrm{c}_3}}^{2}\right)\\
	M_{23} =& I_{3}+l_{\scriptstyle{\mathrm{c}_3}}m_{3}\left(l_{2}c_{3}+l_{\scriptstyle{\mathrm{c}_3}}\right)\\
	M_{33} =& I_{3}+l_{\scriptstyle{\mathrm{c}_3}}^{2}m_{3}.
	\end{align*}The vector of energy contributions due to the acceleration of gravity $ g:=9.81\left[\nicefrac{m}{s^{2}}\right] $ and Coriolis forces $F\left(\theta,\dot{\theta},p\right)\in\mathbb{R}^{3}$ is \begin{align*}
	F\left(\theta,\dot{\theta},p\right) &= \Omega\left(\theta,p\right)\left[\begin{array}{c}
	\dot{\theta}_{1}^{2}\\
	\left(\dot{\theta}_{1}+\dot{\theta}_{2}\right)^{2}\\
	\left(\dot{\theta}_{1}+\dot{\theta}_{2}+\dot{\theta}_{3}\right)^{2}
	\end{array}\right] + g \left[\begin{array}{c}
	k_{1}c_{1}+k_{2}c_{12}+k_{3}c_{123}\\
	k_{2}c_{12}+k_{3}c_{123}\\
	k_{3}c_{123}
	\end{array}\right],
	\end{align*}with $ \Omega \left( \theta, p \right) \in \mathbb{R}^{3 \times 3} $ as\begin{align*}
	\Omega\left(\theta,p\right) &= \left[\begin{array}{ccc}
	l_{1}\left(k_{2}s_{2}+k_{3}s_{23}\right) & -k_{2}l_{1}s_{2}+k_{3}l_{2}s_{3} & -k_{3}\left(l_{1}s_{23}+l_{2}s_{3}\right)\\
	l_{1}\left(k_{2}s_{2}+k_{3}s_{23}\right) & k_{3}l_{2}s_{3} & -k_{3}l_{2}s_{3}\\
	l_{1}k_{3}s_{23} & k_{3}l_{2}s_{3} & 0
	\end{array}\right].
	\end{align*}The generalized force matrix $A_{\tau}\left(\theta,p\right)\in\mathbb{R}^{3\times4}$ is \begin{align*}
	A_{\tau}\left(\theta,p\right) &= \left[\begin{array}{cccc}
	0 & -1 & -l_{1}s_{1}-l_{2}s_{12}-l_{3}s_{123} & l_{1}c_{1}+l_{2}c_{12}+l_{3}c_{123}\\
	0 & -1 & -l_{2}s_{12}-l_{3}s_{123} & l_{2}c_{12}+l_{3}c_{123}\\
	1 & -1 & -l_{3}s_{123} & l_{3}c_{123}
	\end{array}\right].
	\end{align*}
	
	\section{Transformation from $z$ coordinates to $\theta$ coordinates} \label{appendix:z2thetaTransformation}
	
\begin{figure}
		\begin{centering}
			\includegraphics[width=9cm]{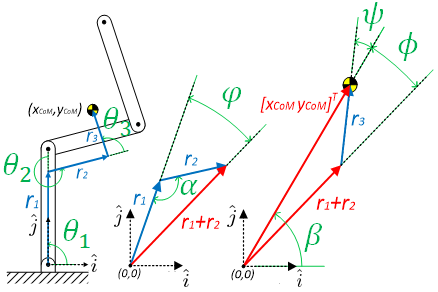}
			\par\end{centering}
		\caption{Geometric representation of vectors and angles used for expressing $\theta_{1}$ and $\theta_{3}$ as a function of $\left[x_{\scriptscriptstyle{\mathrm{CoM}}};y_{\scriptscriptstyle{\mathrm{CoM}}}\right]$ and $\theta_{2}$.\label{fig:Geometry} }
	\end{figure}

	This section reviews the derivation of the transformation from the $z$ to the $\theta$ space, detailed in \cite{Narvaez-Aroche2017}.
	The position of the CoM of the three-link robot can be expressed as a sum of three vectors whose geometric representation is shown in Figure~\ref{fig:Geometry}:\begin{align}
	\left[\begin{array}{c}
	x_{\scriptscriptstyle{\mathrm{CoM}}}\\
	y_{\scriptscriptstyle{\mathrm{CoM}}}
	\end{array}\right] &= k_{0}k_{1}\left[\begin{array}{c}
	c_{1}\\
	s_{1}
	\end{array}\right]+k_{0}k_{2}\left[\begin{array}{c}
	c_{12}\\
	s_{12}
	\end{array}\right]+k_{0}k_{3}\left[\begin{array}{c}
	c_{123}\\
	s_{123}
	\end{array}\right]\nonumber \\
	&= r_{1}+r_{2}+r_{3}.\label{eq:vectors}
	\end{align}
	According to the angles drawn in Figure \ref{fig:Geometry}, we can establish the following relationships: \begin{align}
	\alpha &= \theta_{2}-\pi,\nonumber \\
	\beta &= \arctan\left(\frac{y_{\scriptscriptstyle{\mathrm{CoM}}}}{x_{\scriptscriptstyle{\mathrm{CoM}}}}\right),\label{eq:Beta}\\
	\theta_{1} &= \beta-\phi+\varphi,\label{eq:theta1}\\
	\theta_{3} &= \beta+\psi-\left(\theta_{1}+\theta_{2}\right),\label{eq:theta3}
	\end{align}where for feasible and realistic STS movements $\phi\geq0$, and $\varphi,\psi\in\left[-\nicefrac{\pi}{2},\nicefrac{\pi}{2}\right]$. 
	
	Applying the law of cosines to the triangle formed by vectors $r_{1}+r_{2}$, $r_{3}$, and $\left[x_{\scriptscriptstyle{\mathrm{CoM}}}; y_{\scriptscriptstyle{\mathrm{CoM}}} \right]$, as well as using the trigonometric identity $c_{2}=c_{12}c_{1}+s_{12}s_{1}$, we have\begin{align}
	\phi\left(z\right) &= \arccos\left(\frac{\left\Vert r_{3}\right\Vert ^{2}-\left\Vert r_{1}+r_{2}\right\Vert ^{2}-\left(x_{\scriptscriptstyle{\mathrm{CoM}}}^{2}+y_{\scriptscriptstyle{\mathrm{CoM}}}^{2}\right)}{-2\left\Vert r_{1}+r_{2}\right\Vert \sqrt{x_{\scriptscriptstyle{\mathrm{CoM}}}^{2}+y_{\scriptscriptstyle{\mathrm{CoM}}}^{2}}}\right) \nonumber\\
	&= \arccos\left(\frac{\left(k_{0}k_{3}\right)^{2}-k_{0}^{2}\left(k_{1}^{2}+k_{2}^{2}+2k_{1}k_{2}c_{2}\right)-\left(x_{\scriptscriptstyle{\mathrm{CoM}}}^{2}+y_{\scriptscriptstyle{\mathrm{CoM}}}^{2}\right)}{-2k_{0}\sqrt{k_{1}^{2}+k_{2}^{2}+2k_{1}k_{2}c_{2}}\sqrt{x_{\scriptscriptstyle{\mathrm{CoM}}}^{2}+y_{\scriptscriptstyle{\mathrm{CoM}}}^{2}}}\right).\label{eq:Phi}
	\end{align}
	From the law of sines, for the triangle of vectors $r_{1}$, $r_{2}$, and $r_{1}+r_{2}$, we know \begin{align}
	\varphi\left(z\right) & = \arcsin\left(\frac{\left\Vert r_{2}\right\Vert \sin\alpha}{\left\Vert r_{1}+r_{2}\right\Vert }\right) = \arcsin\left(\frac{k_{2}\sin\left(\theta_{2}-\pi\right)}{\sqrt{k_{1}^{2}+k_{2}^{2}+2k_{1}k_{2}c_{2}}}\right), \label{eq:VarPhi}\\
	\psi\left(z\right) & = \arcsin\left(\frac{\left\Vert r_{1}+r_{2}\right\Vert \sin\phi\left(z\right)}{\left\Vert r_{3}\right\Vert }\right) = \arcsin\left(\frac{\sqrt{k_{1}^{2}+k_{2}^{2}+2k_{1}k_{2}c_{2}}\sin\phi\left(z\right)}{k_{3}}\right).\label{eq:Psi}
	\end{align}
	Plugging \eqref{eq:Beta}, \eqref{eq:Phi}, and \eqref{eq:VarPhi} into \eqref{eq:theta1}, as well as \eqref{eq:Beta}, and \eqref{eq:Psi} into \eqref{eq:theta3}, we define the transformation $h_{1}: z \rightarrow \theta$ as\begin{align*}
	\theta &= \left[\begin{array}{c}
	\arctan\left(\frac{y_{\scriptscriptstyle{\mathrm{CoM}}}}{x_{\scriptscriptstyle{\mathrm{CoM}}}}\right)-\phi\left(z\right)+\varphi\left(z\right)\\
	\theta_{2}\\
	\arctan\left(\frac{y_{\scriptscriptstyle{\mathrm{CoM}}}}{x_{\scriptscriptstyle{\mathrm{CoM}}}}\right)+\psi\left(z\right)-\left(\theta_{1}\left(z\right)+\theta_{2}\right)
	\end{array}\right] =: h_{1}\left(z\right).
	\end{align*}Because this transformation relies on the triangulation of the vectors in equation \eqref{eq:vectors}, it does not hold in the vertical position, where $\theta_{1}=\nicefrac{\pi}{2}$ and $\theta_{2}=\theta_{3}=0$.
	
	The velocity of the CoM of the three-link robot is \begin{align*}
	\left[\begin{array}{c}
	\dot{x}_{\scriptscriptstyle{\mathrm{CoM}}}\\
	\dot{y}_{\scriptscriptstyle{\mathrm{CoM}}}
	\end{array}\right] =& \: k_{0} \dot{\theta}_{2}\left[\begin{array}{c}
	-k_{2}s_{12}-k_{3}s_{123}\\
	k_{2}c_{12}+k_{3}c_{123}
	\end{array}\right] + \left[\begin{array}{cc}
	-y_{\scriptscriptstyle{\mathrm{CoM}}} & -k_{0}k_{3}s_{123}\\
	x_{\scriptscriptstyle{\mathrm{CoM}}} & k_{0}k_{3}c_{123}
	\end{array}\right]\left[\begin{array}{c}
	\dot{\theta}_{1}\\
	\dot{\theta}_{3}
	\end{array}\right].
	\end{align*}
	$ k_{0}k_{3}s_{123} x_{\scriptscriptstyle{\mathrm{CoM}}} - k_{0}k_{3}c_{123} y_{\scriptscriptstyle{\mathrm{CoM}}} = 0$ if $\theta_{1}+\theta_{2}+\theta_{3}=\arctan\left(\frac{y_{\scriptscriptstyle{\mathrm{CoM}}}}{x_{\scriptscriptstyle{\mathrm{CoM}}}}\right)$, and according to expressions \eqref{eq:Beta} and \eqref{eq:theta3} this condition will hold if and only if $\psi=0$; which requires vectors $r_{1}+r_{2}$ and $r_{3}$ to be aligned. In the case of feasible and realistic STS movements, this will only occur in the vertical position. For all other configurations, we calculate the angular velocities of links 1 and 3 as \begin{align*}
	\left[\begin{array}{c}
	\dot{\theta}_{1}\\
	\dot{\theta}_{3}
	\end{array}\right] &= \left[\begin{array}{cc}
	-y_{\scriptscriptstyle{\mathrm{CoM}}} & -k_{0}k_{3}s_{123}\\
	x_{\scriptscriptstyle{\mathrm{CoM}}} & k_{0}k_{3}c_{123}
	\end{array}\right]^{-1} \left(\left[\begin{array}{c}
	\dot{x}_{\scriptscriptstyle{\mathrm{CoM}}}\\
	\dot{y}_{\scriptscriptstyle{\mathrm{CoM}}}
	\end{array}\right]-k_{0}\dot{\theta}_{2}\left[\begin{array}{c}
	-k_{2}s_{12}-k_{3}s_{123}\\
	k_{2}c_{12}+k_{3}c_{123}
	\end{array}\right]\right)\\&=: V\left(z,\dot{z}\right)
	\end{align*}so that the transformation $h_{2}:z,\dot{z}\rightarrow\dot{\theta}$ is defined 
	\begin{align*}
	\dot{\theta} &= \left[\begin{array}{cc}
	1 & 0 \\ 0 & 0 \\ 0 & 1
	\end{array}\right] V \left(z,\dot{z}\right)+\left[\begin{array}{c}
	0\\ \dot{\theta}_{2}\\ 0
	\end{array}\right] =: h_{2} \left(z,\dot{z}\right).
	\end{align*}
	
	The acceleration of the CoM is \begin{align*}
	\left[\begin{array}{c}
	\ddot{x}_{\scriptscriptstyle{\mathrm{CoM}}}\\
	\ddot{y}_{\scriptscriptstyle{\mathrm{CoM}}}
	\end{array}\right] &= a\left(h_{1}\left(z\right),h_{2}\left(z,\dot{z}\right),\ddot{z}\right) + \left[\begin{array}{cc}
	-y_{\scriptscriptstyle{\mathrm{CoM}}} & -k_{0}k_{3}s_{123}\\
	x_{\scriptscriptstyle{\mathrm{CoM}}} & k_{0}k_{3}c_{123}
	\end{array}\right]\left[\begin{array}{c}
	\ddot{\theta}_{1}\\
	\ddot{\theta}_{3}
	\end{array}\right],
	\end{align*}where 
	\begin{align*}
	a\left(\theta,\dot{\theta},\ddot{z}\right):=&-\left[\begin{array}{ccc}
	x_{\scriptscriptstyle{\mathrm{CoM}}} & k_{0}\left(k_{2}c_{12}+k_{3}c_{123}\right) & k_{0}k_{3}c_{123}\\
	y_{\scriptscriptstyle{\mathrm{CoM}}} & k_{0}\left(k_{2}s_{12}+k_{3}s_{123}\right) & k_{0}k_{3}s_{123}
	\end{array}\right]\left[\begin{array}{c}
	\dot{\theta}_{1}^{2}\\
	\dot{\theta}_{2}^{2}\\
	\dot{\theta}_{3}^{2}
	\end{array}\right]-2k_{0}\dot{\theta}_{1}\dot{\theta}_{2}\left[\begin{array}{c}
	k_{2}c_{12}+k_{3}c_{123}\\
	k_{2}s_{12}+k_{3}s_{123}
	\end{array}\right]\\ &-2k_{0}k_{3}\left(\dot{\theta}_{1}+\dot{\theta}_{2}\right)\dot{\theta}_{3}\left[\begin{array}{c}
	c_{123}\\
	s_{123}
	\end{array}\right] + k_{0} \ddot{\theta}_{2} \left[\begin{array}{c}
	-k_{2}s_{12}-k_{3}s_{123}\\
	k_{2}c_{12}+k_{3}c_{123}
	\end{array}\right].
	\end{align*}
	
	Thus\begin{align*}
	\left[\begin{array}{c}
	\ddot{\theta}_{1}\\
	\ddot{\theta}_{3}
	\end{array}\right] & = \left[\begin{array}{cc}
	-y_{\scriptscriptstyle{\mathrm{CoM}}} & -k_{0}k_{3}s_{123}\\
	x_{\scriptscriptstyle{\mathrm{CoM}}} & k_{0}k_{3}c_{123}
	\end{array}\right]^{-1} \left(\left[\begin{array}{c}
	\ddot{x}_{\scriptscriptstyle{\mathrm{CoM}}}\\
	\ddot{y}_{\scriptscriptstyle{\mathrm{CoM}}}
	\end{array}\right]-a\left(h_{1}\left(z\right),h_{2}\left(z,\dot{z}\right),\ddot{z}\right)\right)
	\end{align*}and we can define the transformation $h_{3}:z,\dot{z},\ddot{z}\rightarrow\ddot{\theta}$
	as\begin{align*}
	\ddot{\theta} & = \left[\begin{array}{cc}
	1 & 0\\
	0 & 0\\
	0 & 1
	\end{array}\right]\left[\begin{array}{cc}
	-y_{\scriptscriptstyle{\mathrm{CoM}}} & -k_{0}k_{3}s_{123}\\
	x_{\scriptscriptstyle{\mathrm{CoM}}} & k_{0}k_{3}c_{123}
	\end{array}\right]^{-1} \left(\left[\begin{array}{c}
	\ddot{x}_{\scriptscriptstyle{\mathrm{CoM}}}\\
	\ddot{y}_{\scriptscriptstyle{\mathrm{CoM}}}
	\end{array}\right]-a\left(h_{1}\left(z\right),h_{2}\left(z,\dot{z}\right),\ddot{z}\right)\right) +\left[\begin{array}{c}
	0 \\ \ddot{\theta}_{2} \\ 0\end{array}\right] =: h_{3}\left(z,\dot{z},\ddot{z}\right).
	\end{align*}
	
	For compactness, we use 
	\begin{align}
	h\left(z,\dot{z},\ddot{z}\right)&:=\left[
	h_{1}\left(z\right); \;
	h_{2}\left(z,\dot{z}\right); \;
	h_{3}\left(z,\dot{z},\ddot{z}\right)\right]\label{eq:KinematicsTransform}
	\end{align}to denote the transformation from $z$ coordinates to $\theta$ coordinates.
	
	\section{Kinematics of the Center of Mass} \label{appendix:CoMKinematics}
	
	The position and velocity coordinates of the CoM of the three-link planar robot in Figure~\ref{fig:Robot} are computed from the kinematic equations derived in~\cite{Narvaez-Aroche2017}, with the mapping $\zeta:\mathbb{R}^6\times\mathbb{R}^{12}\rightarrow\mathbb{R}^4$ defined as:\begin{align*}
	\left[ \begin{array}{c}
	x_{\scriptscriptstyle{\mathrm{CoM}}}\\ 
	y_{\scriptscriptstyle{\mathrm{CoM}}}\\ 
	\dot x_{\scriptscriptstyle{\mathrm{CoM}}}\\ 
	\dot y_{\scriptscriptstyle{\mathrm{CoM}}}\\
	\end{array} \right] &= \left[\begin{array}{c}
	k_{0}\left(k_{1}c_{1}+k_{2}c_{12}+k_{3}c_{123}\right)\\
	k_{0}\left(k_{1}s_{1}+k_{2}s_{12}+k_{3}s_{123}\right)\\
	-\dot{\theta}_{1}y_{\scriptscriptstyle{\mathrm{CoM}}}-\dot{\theta}_{2}k_{0}\left(k_{2}s_{12}+k_{3}s_{123}\right)-\dot{\theta}_{3}k_{0}k_{3}s_{123}\\
	\dot{\theta}_{1}x_{\scriptscriptstyle{\mathrm{CoM}}}+\dot{\theta}_{2}k_{0}\left(k_{2}c_{12}+k_{3}c_{123}\right)+\dot{\theta}_{3}k_{0}k_{3}c_{123}
	\end{array}\right] =: \zeta\left(x,p\right). 
	\end{align*}
	\section*{Acknowledgments}
	
	The first author would like to thank the Consejo Nacional de Ciencia y Tecnolog\'{i}a (CONACYT), the Fulbright-Garc\'{i}a Robles program, and the University of California Institute for Mexico and the United States (UC MEXUS) for the scholarships that have made possible his Ph.D. studies, as well as Dr. Maria Vrakopoulou for the drawings in Figure \ref{fig:STSPhases}, and Matthias Hirche for his comments on the manuscript. The authors gratefully acknowledge support from the National Science Foundation under grant ECCS-1405413. Andrew Packard also acknowledges the generous support of the FANUC Corporation.
	
	\bibliographystyle{IEEEtran}
	\bibliography{IEEEabrv,STSBiblio}
	
\end{document}